\documentclass[12pt]{article}
\usepackage{epsfig}
\begin{document}
%%%%%%%%%%%%%%%%%%%%%%%%%%%%%%%%%%%%%%%%%%%%%%%%%%%%%%%%%%%%%%%%%%%%%%%%%
\title{Medium effects in antikaon-induced $\Xi^-$ hyperon production
 on nuclei near threshold}
\author{E. Ya. Paryev$^{1,2}$\\
{\it $^1$Institute for Nuclear Research, Russian Academy of Sciences,}\\
{\it Moscow 117312, Russia}\\
{\it $^2$Institute for Theoretical and Experimental Physics,}\\
{\it Moscow 117218, Russia}}
%%==============================================================
%%==============================================================

\renewcommand{\today}{}
\maketitle

\begin{abstract}
We study the antikaon-induced inclusive cascade $\Xi^-$ hyperon
production from $^{12}$C and $^{184}$W target nuclei near threshold within a nuclear spectral function
approach. The approach describes incoherent direct $\Xi^-$ hyperon production in elementary
${K^-}p \to {K^+}\Xi^-$ and ${K^-}n \to {K^0}\Xi^-$ processes as well as takes into account the influence
of the scalar nuclear $K^-$, $K^+$, $K^0$, $\Xi^-$ and their Coulomb potentials on these processes.
We calculate the absolute differential and total cross sections for the production of $\Xi^-$ hyperons
off these nuclei at laboratory angles $\le$ 45$^{\circ}$ by $K^-$ mesons with momenta of 1.0 and 1.3 GeV/c,
which are close to the threshold momentum (1.05 GeV/c) for $\Xi^-$ hyperon production on the free
target nucleon at rest. We also calculate the momentum dependence of the transparency ratio for the
$^{184}$W/$^{12}$C combination for $\Xi^-$ hyperons at these $K^-$ beam momenta.
We show that the $\Xi^-$ differential and total (absolute and relative) production cross sections
at the considered initial momenta reveal a distinct sensitivity to the variations
in the scalar $\Xi^-$ nuclear potential at saturation density $\rho_0$, studied in the paper,
in the low-momentum region of 0.1--0.6 GeV/c.
We also demonstrate that for the subthreshold $K^-$ meson momentum of 1.0 GeV/c there is, contrary to the
case of its above threshold momentum of 1.3 GeV/c, a strong sensitivity of the transparency ratio
for $\Xi^-$ hyperons to the considered changes in the $\Xi^-$ nuclear potential at all outgoing $\Xi^-$
momenta as well, which cannot be masked, as in the case of differential and total observables mentioned above,
by that associated with the possible changes in the poorly known experimentally ${\Xi^-}N$ inelastic cross section. Therefore, the measurement of these absolute and relative observables in a dedicated experiment at
the J-PARC Hadron Experimental Facility will provide valuable information on the $\Xi^-$ in-medium properties,
which will be complementary to that deduced from the study of the inclusive ($K^-$,$K^+$) reactions
at incident momenta of 1.6--1.8 GeV/c in the $\Xi^-$ bound and quasi-free regions.
\end{abstract}

\newpage

\section*{1 Introduction}

\hspace{0.5cm} Essential progress has been made over the last decades in studying the properties
of the light $K$ and ${\bar K}$ [1--3], heavy $K^*(892)$ and ${\bar K}^*(892)$ [4--10], $K_1(1270)$ [11--15]
mesons as well as of the $\Lambda$, $\Sigma$, $\Lambda(1405)$, $\Sigma(1385)$, $\Lambda(1520)$
[16--26] hyperons with strangeness $S=-1$ in nuclear matter. The knowledge of these properties is important
for understanding both the size and stability of neutron stars, where kaonic and hyperonic matters are
expected to appear at their cores, and a partial restoration of chiral symmetry in a dense nuclear medium.
What concerns the $S=-2$ sector, the situation, in particular, with the ${\Xi^-}N$ and ${\Xi^-}$--nucleus interactions
\footnote{$^)$The baryon--baryon interactions of the type ${\Lambda}{\Lambda}$, ${\Lambda}{\Sigma}$ and
${\Sigma}{\Sigma}$ are also considered in the strangeness $S=-2$ sector.}$^)$
is much less conclusive because of the scarcity of the respective $\Xi^-$ scattering, production and hypernuclear data. The experimental knowledge on these interactions is poor currently.
The study of $\Xi^-$ hypernuclei and the momentum correlations of $p$ and $\Xi^-$, produced in relativistic
heavy ion collisions, as well as the first principle lattice HAL QCD calculations for masses very close to
the physical point provide a valuable information on them at low energies (see, for example, [2, 26, 27, 28]).
At present, there are only a few sets of experimental data on the $\Xi^-$ hyperon properties in the nuclear medium.
Phenomenological information deduced in Refs. [29, 30] from old emulsion data, from missing-mass
measurements [31--34] in the inclusive ($K^-$,$K^+$) reaction on nuclear targets at incident momenta of
1.6--1.8 GeV/c in the $\Xi^-$ bound and quasi-free regions with insufficiently good (10 MeV or worse) [31, 32, 33]
and moderate (5.4 MeV) [34] energy resolutions, from the analysis the results of these measurements
yet in Refs. [35, 36] using Green's function method of the DWIA as well as the theoretical predictions [37--41]
indicate that the $\Xi^-$ cascade hyperon feels only a weak attractive potential in nuclei,
the depth of which is not so large, $\sim$ -(10--20) MeV at central nuclear density and at rest
\footnote{$^)$The depth of -14 MeV for the ${\Xi^-}$--nucleus potential in the nuclear interior is considered
in the literature as the canonical or as the benchmark one [26].}$^)$
.
Moreover the new recent emulsion KEK-PS E373 [42] and J-PARC E07 [43] experiments,
in which the remarkable events named, respectively, "KISO" and "IBUKI" were observed, reported the evidence
of a likely the Coulomb-assisted nuclear 1$p$ single-particle $\Xi^-$ state of the bound
$\Xi^-$--$^{14}$N(g.s.) system with binding energy of about 1 MeV, which also testifies in favor of shallow
$\Xi^-$--nucleus potential. The observation of two another hypernuclear events, "KINKA" and "IRRAWADDY",
in the E373 and E07 experiments, respectively, gives the first indication of the nuclear 1$s$ $\Xi^-$ state
of an extremely deep $\Xi^-$--$^{14}$N bound system [44], which also suggests an attractive low-energy potential
of the $\Xi^-$ hyperon in nuclear matter. In view of these observations, it is interesting to note that (I) the
possibility of the existence of such exotic $\Xi$ hypernuclei as the lightest three-body ${\Xi}NN$ and
four-body ${\Xi}NNN$ bound systems has been investigated in Refs. [45] and [46] using, respectively, a
Gaussian Expansion Method with two modern ${\Xi}N$ interactions and a pionless halo effective field theory
and that (II) the ground state energies of some another exotic $\Xi^-$ hypernuclei -- double $\Xi^-$ hypernuclei
have been calculated in Ref. [47] within the respective three-body model.
To improve essentially our understanding of the $\Xi^-N$ interaction the high-quality data on the doubly
strange $S=-2$ hypernuclei are needed. It is expected that in the near future high resolution (up to 1.4 MeV)
and high statistics data on the missing-mass spectra for the $^{12}$C($K^-$,$K^+$) reaction around the $\Xi^-$
production threshold and the X-ray data on the level shifts and width broadening of the $\Xi^-$ atomic states in the
$\Xi^-$ atoms will become available from the planned dedicated E70 and E03 experiments at the J-PARC Hadron
Experimental Facility [48].

 With regard to the $\Xi^-$ production data and in addition to the aforesaid experimental activities,
the $\Xi^-$ yield has been measured at various high beam energies in central Au+Au and Pb+Pb collisions
at LHC [49], RHIC [50, 51, 52], SPS [53, 54, 55] and AGS [56]. At SIS energies, the first results on a deep
subthreshold production of $\Xi^-$ hyperons in Ar+KCl reactions at beam kinetic energy of 1.76A GeV have been
reported by the HADES Collaboration [57]. A surprisingly high $\Xi^-$ yield over that for $\Lambda$ hyperons
(the abundance ratio) has been observed in this experiment, which exceeds an earlier predictions of a
statistical model [58] and a relativistic transport approach [59]. Later on, the $\Xi^-/\Lambda$ abundance
ratio measured by HADES is found to be essentially consistent with the result of calculation within the
Relativistic Vlasov-Uehling-Uhlenbeck (RVUU) transport model accounting for the contributions of
hyperon--hyperon scattering channels $YY \leftrightarrow {\Xi}N$ $(Y=\Lambda, \Sigma$),
which enhance strongly the yield of $\Xi^-$ [60]. On the other hand, the UrQMD transport model [61],
using the $YY$ cross sections provided in [60], could not satisfactorily explain the measured in [57]
$\Xi^-/\Lambda$ ratio. The possible reasons of the discrepancy between the results of Refs. [60] and [61]
are discussed in [61]. Also the data on $\Xi^-$ production in $p$+$p$ and $p$+Be, $p$+Pb collisions have
been collected, respectively, in the NA61/SHINE [62] and NA57 [63] experiments performed
at the SPS at a beam momentum of 158 GeV/c. Recently, the subthreshold $\Xi^-$ production in collisions
of $p$(3.5 GeV)+Nb has been observed for the first time by the HADES Collaboration as well [64]
\footnote{$^)$The kinetic threshold energy for $\Xi^-$ production in free elementary nucleon--nucleon
collisions is 3.74 GeV.}$^)$
.
It was shown that, in spite of the fact that proton-induced reactions have much simpler dynamics compared
to the case of nucleus-nucleus interactions,
the available statistical model and two transport approaches (UrQMD and GIBUU)
predictions turned out to be substantially lower than the measured in these collisions $\Xi^-$ yield.
Contrary to the case of deep subthreshold $\Xi^-$ hyperon production in Ar+KCl reactions
at beam kinetic energy of 1.76A GeV [60], the strangeness-exchange reactions $YY \leftrightarrow {\Xi}N$
are found to be of minor importance for subthreshold $\Xi^-$ creation in $pA$ interactions. This means that
a new $\Xi^-$ puzzle appears in these interactions. Possible explanation for it, like in-medium modifications
of kaon and $\Xi^-$ hyperon properties, should be studied. It is worth noting that the
exclusive photoproduction of $\Xi^-$ and $\Xi^0$ hyperons in reactions ${\gamma}p \to K^+{K^+}{\Xi^-}$ and
${\gamma}p \to K^+{K^+}{\pi^-}{\Xi^0}$ has been investigated at JLab for near-threshold photon energies
using the CLAS detector [65, 66, 67]. Finally, it should be also pointed out that the study of production of
ground and excited $\Xi$ baryons is planned to be performed in the future PANDA experiment with an antiproton
beam at FAIR [68] and with the upgraded HADES detector in $p$+$p$ reactions at a beam kinetic energy
of 4.5 GeV [69].

In addition to the aforementioned ($K^-$,$K^+$) reactions on nuclear targets at beam momenta
of 1.6--1.8 GeV/c, the medium modification of the $\Xi^-$ hyperon could be probed directly through the
another inclusive near-threshold ($K^-$,$\Xi^-$) reactions
\footnote{$^)$Their threshold momentum for free $K^-N$ collisions is about 1.05 GeV/c.}$^)$
on nuclei at J-PARC. The advantage of the latter reactions compared to the former ones is that
they have a much cleaner dynamics due to the fact that in them the role played by the
two-step processes is expected to be smaller [70].
As a guidance for such future measurements, herein we give the predictions
for the absolute differential and total cross sections for near-threshold production of $\Xi^-$ hyperons in
${K^-}^{12}{\rm C} \to {\Xi^-}X$ and ${K^-}^{184}{\rm W} \to {\Xi^-}X$ reactions at laboratory angles of
$\le$ 45$^{\circ}$ by incident $K^-$ mesons with momenta of 1.0 and 1.3 GeV/c as well as for their relative yields
from these reactions within three different scenarios (see below) for the effective scalar nuclear potential that
$\Xi^-$ hyperon feels in the medium.
The present calculations are based on a first-collision model, developed in Ref. [71] for the
description of the inclusive $\phi$ meson production on nuclei in near-threshold pion-induced reactions
and extended to account for these scenarios for the $\Xi^-$ nuclear potential.
Comparison the results of our present calculations with the respective data, which could be
taken in future dedicated experiment using $K^-$ beams at the J-PARC Hadron Experimental Facility,
will provide a deeper insight into the $\Xi^-$ in-medium properties and information obtained from this
comparison will supplement that deduced from the ($K^-$,$K^+$) reactions on nuclear targets.

\section*{2 Model: direct processes of $\Xi^-$ hyperon production in nuclei}

\hspace{0.5cm} Direct production of $\Xi^-$ hyperons in antikaon--nucleus
reactions at near-threshold incident $K^-$ momenta below 1.3 GeV/c can occur
in the following $K^-p$ and $K^-n$ elementary processes, which have
the lowest free production threshold momentum ($\approx$ 1.05 GeV/c):
%formula(1)
\begin{equation}
K^-+p \to K^++\Xi^-,
\end{equation}
%formula(2)
\begin{equation}
K^-+n \to K^0+\Xi^-.
\end{equation}
We can neglect the contribution to the $\Xi^-$ yield from the processes $K^-N \to K{\Xi^-}{\pi}$
with one pion in the final state at initial momenta of interest due to larger their thresholds
($\approx$ 1.34 GeV/c) in free $K^-N$ interactions.

Following [72], for numerical simplicity we will include the in-medium modification of the initial $K^-$
meson as well as of the final $K^+$, $K^0$ mesons and $\Xi^-$ hyperons,
involved in the production processes (1), (2),
in terms of their average in-medium masses $<m_{K^-}^*>$, $<m_{K^+}^*>$, $<m_{K^0}^*>$ and $<m_{\Xi^-}^*>$
instead of their local effective mass $m_{K^-}^*(|{\bf r}|)$, $m_{K^+}^*(|{\bf r}|)$, $m_{K^0}^*(|{\bf r}|)$
and $m_{\Xi^-}^*(|{\bf r}|)$ in the in-medium cross sections
of these processes, with average in-medium masses defined as:
%formula(3)
\begin{equation}
<m^*_{h}>=m_{h}+U_h\frac{<{\rho_N}>}{{\rho_0}}+V_{{\rm c}h}(R_{\rm c}),
\end{equation}
where $h$ stands for $K^-$, $K^+$, $K^0$ and $\Xi^-$.
Here, $m_{h}$ is the hadron mass in the free space, $U_h$ is the value of its effective scalar
nuclear potential (or its in-medium strong interaction mass shift) at normal nuclear matter
density ${\rho_0}$, $<{\rho_N}>$ is the average nucleon density and $V_{{\rm c}h}(R_{\rm c})$
is the charged hadron Coulomb potential
\footnote{$^)$This potential for positively and negatively charged hadrons amounts approximately
to +3.1 and +15.4 MeV and -3.1 and -15.4 MeV for $^{12}$C$_{6}$ and $^{184}$W$_{74}$ nuclei,
respectively [73, 74].}$^)$
of uniform target nucleus charge distribution with a radius
of $R_{\rm c}=1.22$A$^{1/3}$ fm taken at the point $|{\bf r}|=R_{\rm c}$
\footnote{$^)$Since the incident $K^-$ mesons are absorbed to a large extent on the surface of the
nucleus due to their strong initial-state interactions.}$^)$
.
In the present work, for nuclei $^{12}$C and $^{184}$W of interest,  the ratio $<{\rho_N}>/{\rho_0}$,
was chosen as 0.55 and 0.76, respectively. For the $K^+$ mass shift $U_{K^+}$
we will adopt the following momentum-independent option: $U_{K^+}=+22$ MeV [75, 76].
Similar to $K^+$ ($K^+=|{\bar s}u>$),
the $K^0$ meson optical potential is expected to be repulsive as well due to its quark content
($K^0=|{\bar s}d>$) and should be close to that for $K^+$ in symmetric nuclear matter with equal
densities of protons and neutrons [76, 77]. Therefore, the same option will be employed
for the $K^0$ mass shifts $U_{K^0}$ in the case of $^{12}$C target nucleus,
having $Z=6$ protons and $N=6$ neutrons, i.e.: $U_{K^0}=U_{K^+}=+22$ MeV [76, 77, 78].
On the other hand, the $K^+$ and $K^0$ mesons are not mass degenerate
in an asymmetric nuclear medium [76, 77, 78]. Following Refs. [76, 77], in the subsequent study for the $K^0$
effective scalar potential depth $U_{K^0}$ we will use the following scenario: $U_{K^0}=+40$ MeV in
the case of $^{184}$W nucleus, containing $Z=74$ protons and $N=110$ neutrons. The $K^-$ nuclear potential
at saturation density $U_{K^-}$, relevant for the incoming $K^-$ meson momentum range of 1.0--1.3 GeV/c,
can be chosen as $U_{K^-}=-40$ MeV according to the dispersion analysis performed in Ref. [79].

Let us now determine the $\Xi^-$ hyperon effective scalar potential $U_{\Xi^-}$, entering into Eq. (3) and
appropriate for our present study. In line with the above-mentioned, a nuclear mean-field potential
$U_{\Xi^-}$, acting on a low-momentum $\Xi^-$ hyperon embedded in nuclear matter, is a moderately
attractive and could be in the vicinity of -14 MeV at saturation density $\rho_0$.
Since the accessible range of the $\Xi^-$
hyperon vacuum momenta at incident beam momenta of interest is about of 0.2--0.8 GeV/c (see below),
it is helpful to estimate this potential in symmetric nuclear matter also for such in-medium $\Xi^-$ momenta
at density $\rho_0$. We will rely on the constituent quark model, according to which the quark structure
of the $\Xi^-$ hyperon is $\Xi^-$=$|dss>$. Therefore, the $\Xi^-$ mean-field scalar
$U_{S{\Xi^-}}$ and vector $U_{V{\Xi^-}}$ potentials are about 1/3 of those
$U_{SN}$ and $U_{VN}$ of a nucleon [80, 81] when in-medium nucleon and $\Xi^-$ hyperon velocities
$v^{\prime}_N$ and $v^{\prime}_{\Xi^-}$ relative to the nuclear matter are equal to each other, i.e.,
%formula(4)
$$
U_{S{\Xi^-}}(v^{\prime}_{\Xi^-},\rho_N)=\frac{1}{3}U_{SN}(v^{\prime}_{N},\rho_N),
$$
\begin{equation}
U_{V{\Xi^-}}(v^{\prime}_{\Xi^-},\rho_N)=\frac{1}{3}{\alpha}U_{VN}(v^{\prime}_{N},\rho_N);\,\,\,
v^{\prime}_{N}=v^{\prime}_{\Xi^-}.
\end{equation}
The latter term in Eq. (4) corresponds, to the following relation between
the respective in-medium nucleon momentum $p^{\prime}_N$ and the $\Xi^-$ one $p^{\prime}_{\Xi^-}$:
%formula(5)
\begin{equation}
p^{\prime}_{N}=\frac{<m^*_{N}>}{<m^*_{\Xi^-}>}p^{\prime}_{\Xi^-}.
\end{equation}
However, for reasons of numerical simplicity, calculating the $\Xi^-$--nucleus single-particle potential
(or the so-called Schr${\ddot{\rm o}}$dinger equivalent potential $V_{{\Xi^-}A}^{\rm SEP}$), we will
employ in expression (5) free space nucleon and $\Xi^-$ hyperon masses $m_N$ and $m_{\Xi^-}$ instead of
their average in-medium masses $<m^*_{N}>$ and $<m^*_{\Xi^-}>$.
Then, this potential $V_{{\Xi^-}A}^{\rm SEP}$ can be
defined as [80, 81]
\footnote{$^)$The space-like component of the $\Xi^-$ vector self-energy is ignored here.}$^)$
:
%formula(6)
\begin{equation}
V_{{\Xi^-}A}^{\rm SEP}(p^{\prime}_{\Xi^-},\rho_N)=
\sqrt{\left[m_{\Xi^-}+U_{S{\Xi^-}}(p^{\prime}_{\Xi^-},\rho_N)\right]^2+({p^{\prime}_{\Xi^-}})^2}
+U_{V{\Xi^-}}(p^{\prime}_{\Xi^-},\rho_N)-\sqrt{m_{\Xi^-}^2+({p^{\prime}_{\Xi^-}})^2}.
\end{equation}
The relation between the potentials $U_{\Xi^-}$ and $V_{{\Xi^-}A}^{\rm SEP}$ at normal nuclear
matter density is given by
%formula(7)
\begin{equation}
U_{\Xi^-}({p^{\prime}_{\Xi^-}})=
\frac{\sqrt{m^2_{\Xi^-}+({p^{\prime}_{\Xi^-}})^2}}{m_{\Xi^-}}V_{{\Xi^-}A}^{\rm SEP}({p^{\prime}_{\Xi^-}}).
\end{equation}
Employing the momentum-dependent parametrization for the nucleon scalar and vector potentials
at saturation density $\rho_0$ from [82]
%formula(8)
\begin{equation}
U_{SN}(p^{\prime}_{N},\rho_0)=-\frac{494.2272}{1+0.3426\sqrt{p^{\prime}_{N}/p_F}}~{\rm MeV},
\end{equation}
%formula(9)
\begin{equation}
U_{VN}(p^{\prime}_{N},\rho_0)=\frac{420.5226}{1+0.4585\sqrt{p^{\prime}_{N}/p_F}}~{\rm MeV}
\end{equation}
(where $p_F=1.35$ fm$^{-1}=$0.2673 GeV/c)
and using Eqs. (4)--(6), we calculated the momentum dependence of potential
$V_{{\Xi^-}A}^{\rm SEP}$ at density $\rho_0$.
In doing so,
we have also made an adjustment by multiplying the vector nucleon potential in Eq. (4) by a factor $\alpha$
of $\alpha=$1.068 [78] to get for $\Lambda$ hyperon potential $V_{{\Lambda}A}^{\rm SEP}$ at zero momentum
and at density $\rho_0$ a value consistent with the
experimental one of -(32$\pm$2) MeV, extracted from data on binding energies
of $\Lambda$ single-particle states in nuclei [83]. The adjusted in such manner $\Xi^-$ potential
$V_{{\Xi^-}A}^{\rm SEP}$ is shown in figure 1 by solid curve.
One can see that in this case the $\Xi^-$--nucleus potential
is attractive for all momenta $\le$ 0.8 GeV/c with the value of
$V_{{\Xi^-}A}^{\rm SEP}(0)=U_{\Xi^-}(0)\approx$-15 MeV,
whereas it will be repulsive for higher momenta.
%%%%%%%%%%%%%%%%%%%%%%%%%%%%%%%%%%%%%%%%%%%%%%%%%%%%%%%%%%%%%%%%%%%%%%%%%%%%%%%%%%%%%%%%%%%%%
\begin{figure}[!ht]
\begin{center}
\includegraphics[width=18.0cm]{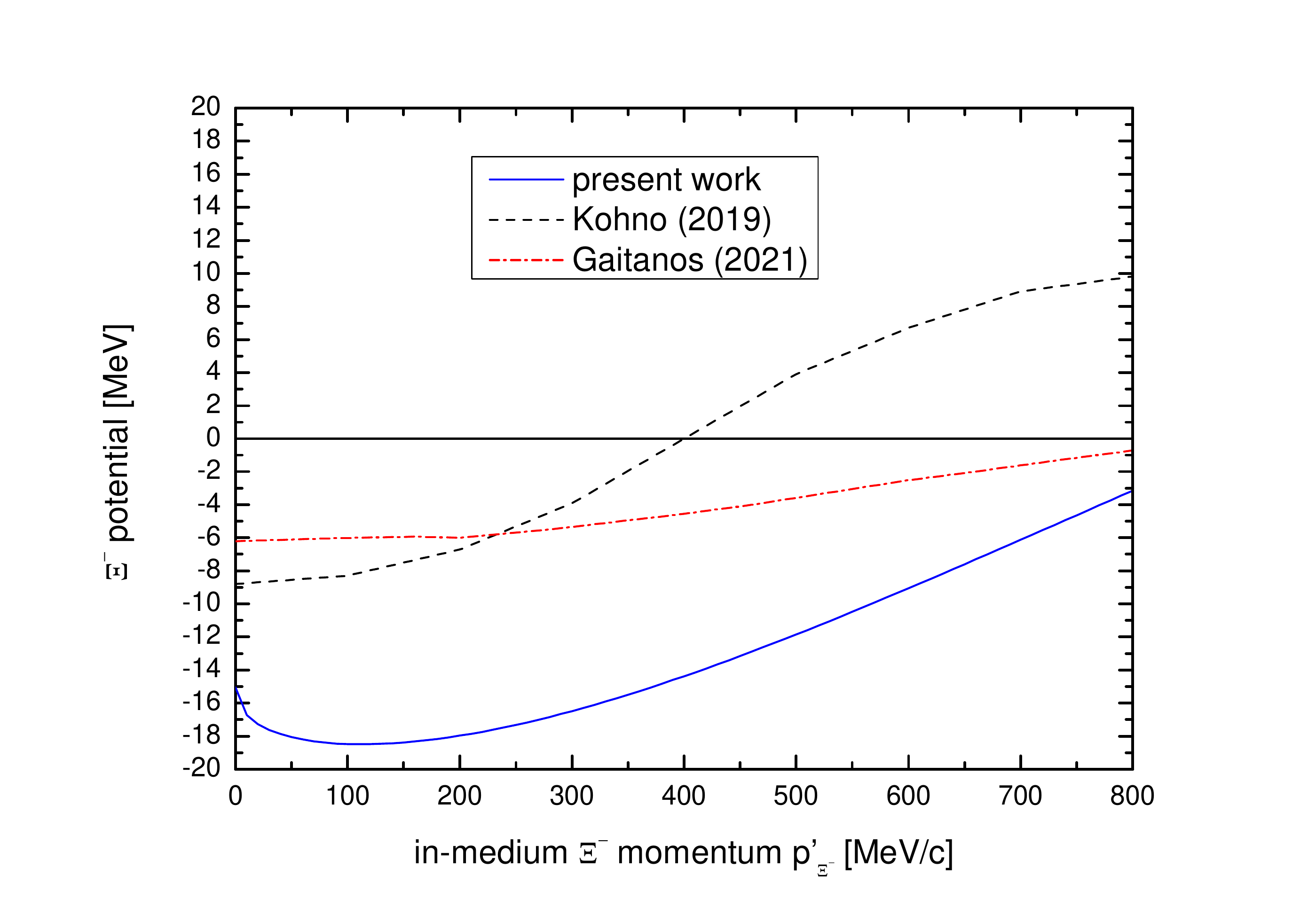}
\vspace*{-2mm} \caption{(Color online) Momentum dependence of the Schr$\ddot{\rm o}$dinger equivalent
${\Xi^-}$ hyperon potential at density $\rho_0$. For notation see the text.}
\label{void}
\end{center}
\end{figure}
%%%%%%%%%%%%%%%%%%%%%%%%%%%%%%%%%%%%%%%%%%%%%%%%%%%%%%%%%%%%%%%%%%%%%%%%%%%%%%%%%%%%%%%%%%%%%
The above value is well consistent with the $\Xi^-$--nucleus Woods-Saxon potential depths of about
-16 and -14 MeV, deduced from the analysis of the data in the $\Xi^-$-bound state region of the missing-mass
spectra for $^{12}$C($K^-$,$K^+$) reaction taken by the KEK--PS E224 [31] and BNL--AGS E885 [32] Collaborations
at incident momenta of 1.6 and 1.8 GeV/c, respectively. Moreover, it is also in good agreement with that of
$\approx$-16 MeV for the $\Xi^-$ in-medium mass shift predicted very recently in Ref. [84] within the framework
of the in-medium modified chiral soliton model. In addition, results for two $\Xi^-$ single-particle potentials
for symmetric nuclear matter from the recent literature are shown in Fig. 1 as well, namely: the chiral effective
field theory (EFT) at the next-to-leading order (NLO) based potential [74] (dashed curve) and the
NLD potential [85] (dotted-dashed curve). It is seen that the latter potential is more weakly attractive at all
momenta considered than that, calculated in the present work. Its momentum dependence is a relatively weak with
the value of about -6 MeV at zero momentum with respect to the surrounding nuclear matter. It is worth
noting that this potential is compatible [85] with recently updated lattice QCD calculations [86] (cf. [87]),
which show a slightly more shallow $\Xi^-$ single-particle potential in the nuclear medium at the density
$\rho_0$ for momenta below $\approx$ 600 MeV/c with the central value of -4 MeV with the statistical error
about $\pm$2 MeV at zero momentum and rather moderate repulsive potential at higher momenta.
It should be pointed out that
this lattice QCD based potential at zero momentum relative to the bulk matter is consistent with the
corresponding values of -3.8 to -5.5 MeV predicted in Ref. [88] within the updated chiral
EFT at the NLO (cf. [89]). An essential difference between present, NLD
potentials and that from Ref. [74] is that the latter shows a much stronger momentum dependence and turns from
attractive with the value of $V_{{\Xi^-}A}^{\rm SEP}(0)\approx$-9 MeV, which is also in line with that of -7 MeV
predicted in Ref. [90] for the Nijmegen ESC08c potential, to repulsive at about 400 MeV/c momentum
and reaches the value $\approx$+10 MeV at $\Xi^-$ momentum of about 800 MeV/c. On the other hand, contrary to
the above, the in-medium calculations [91] and [92], based on the baryon-baryon interactions in the $S=-2$ sector
derived using the chiral EFT in the leading order (LO), yield $V_{{\Xi^-}A}^{\rm SEP}(0)\approx$+6 MeV [91]
and $\approx$+9 MeV [88, 92] at normal nuclear density. So, nowadays there are considerable
uncertainties in determining the $\Xi^-$ nuclear potential both at threshold and at finite momenta below 1 GeV/c.
These uncertainties
could be essentially reduced, comparing, for example, the results of the present model calculations with the
respective precise experimental data on direct $\Xi^-$ production on nuclei. These data could be obtained in
future dedicated experiment at J-PARC using near-thresholds $K^-$ beams. Here, the $\Xi^-$ hyperons could be
registered via the hadronic decay chain $\Xi^- \to {\Lambda}\pi^- \to p{\pi^-}{\pi^-}$, which has a total branching
ratio of about 64\%. It should be noted that a similar comparison of a model cross sections with data on
$\eta^{\prime}$ meson photoproduction off carbon and niobium target nuclei was adopted in Refs. [93, 94] to
extract its effective scalar potential in the nuclear medium. It should be noted that some hint on the possible strength
of the $\Xi^-$ optical potential for $\Xi^-$--$^8$Li system at momenta of interest comes from the recent
analysis [95] of the BNL--E906 measurement of the spectrum of the $^9$Be($K^-$,$K^+$) reaction at incident $K^-$
momentum of 1.8 GeV/c and $K^+$ forward-direction lab angles of $\theta_{K^+}=$1.5$^{\circ}$--8.5$^{\circ}$
in the $\Xi^-$ quasi-free region within the DWIA using the optimal Fermi-averaged $K^-p \to K^+\Xi^-$ amplitude.
The strength of (-17$\pm$6) MeV for this potential in the Woods-Saxon prescription was deduced from this analysis
in the momentum transfer region $q$ $\simeq$ 390--600 MeV/c [95]. If we assume that $p^{\prime}_{\Xi^-}=q$, then this
value can be to some extent interpreted as $\Xi^-$--nucleus potential well depth in the momentum range of
$\simeq$ 390--600 MeV/c and, therefore, can be compared with the predictions presented in Fig. 1. It can be seen
that only our calculation, which gives $V_{{\Xi^-}A}^{\rm SEP}(390~{\rm MeV/c})\approx$-14 MeV and
$V_{{\Xi^-}A}^{\rm SEP}(600~{\rm MeV/c})\approx$-9 MeV, is compatible with the result of [95] over the given
momentum range. Accounting for the relation (7) between the potentials $U_{\Xi^-}$ and $V_{{\Xi^-}A}^{\rm SEP}$
at normal nuclear matter density and using our results shown in Fig. 1, we can readily get that the former one
varies from $\approx$ -18 MeV to -10 MeV when $\Xi^-$ momentum changes from $\sim$ 200 MeV/c to 600 MeV/c with
average value of $\approx$ -14 MeV, which is quite consistent with that inferred by Khaustov {\it et al.} [32]
at zero momentum. Therefore, it is natural to employ for the quantity $U_{\Xi^-}$ in our work the canonical
value of $U_{\Xi^-}=-14$ MeV in whole $\Xi^-$ momentum range studied (see, also [96, 97, 98]).
On the other hand, in this momentum range this potential, as follows from Fig. 1 and Eq. (7), might be repulsive
at $\Xi^-$ hyperon momenta $>$ 600 MeV/c reaching the value $\sim$ +14 MeV at momentum $\sim$ 1.0 GeV/c.
Therefore, to account for all the possible variations in the $\Xi^-$ nuclear potential at all outgoing $\Xi^-$
momenta, we will also both ignore it
\footnote{$^)$Which corresponds to almost zero $\Xi^-$ potential at finite momenta of interest, predicted in
particular in Ref. [86].}$^)$
in our
calculations and use for it in them the value of +14 MeV. Thus, the results will be presented with the three basic
cases of a $\Xi^-$ potential: i) $U_{\Xi^-}=-14$ MeV, ii) $U_{\Xi^-}=0$ MeV and iii) $U_{\Xi^-}=+14$ MeV (cf. [74])
at finite momenta, accessible in calculation of the $\Xi^-$ production in $K^-A$ reactions at beam momenta of
interest for both target nuclei of $^{12}$C and $^{184}$W
\footnote{$^)$It should be noticed that this assumes a purely isoscalar $\Xi^-$ potential
$U_{\Xi^-}^{\rm isoscalar}(r)$, associated with the total nuclear density $\rho_n(r)+\rho_p(r)$, for both
these targets and does not account for the contribution to the total strong interaction $\Xi^-$ potential
$U_{\Xi^-}(r)$ from the isovector one $U_{\Xi^-}^{\rm isovector}(r)$, appearing in the neutron-rich nuclei
due to the neutron excess density $\rho_n(r)-\rho_p(r)$. The latter might be nonnegligible in the $^{184}$W
nucleus. Let us estimate it. According to [2], these low-energy potentials can be expressed as:
$U_{\Xi^-}^{\rm isoscalar}(r)=\alpha_0[\rho_n(r)+\rho_p(r)]$ and
$U_{\Xi^-}^{\rm isovector}(r)=\alpha_1[\rho_n(r)-\rho_p(r)]$. Assuming that $\rho_{n(p)}(r)=N(Z)\rho(r)$, we
get that the ratio $U_{\Xi^-}^{\rm isovector}(r)/U_{\Xi^-}^{\rm isoscalar}(r)$ is equal to
$[(N-Z)/A][\alpha_1/\alpha_0]$. The ratio $\alpha_1/\alpha_0$ can be evaluated as follows. Take for example
the values $U_{\Xi^-}(0)=U_{\Xi^-}^{\rm isoscalar}(0)=2\alpha_0\rho_n(0)=-4$ MeV in symmetric nuclear matter
and $U_{\Xi^-}(0)=2\alpha_0\rho_n(0)+2\alpha_1\rho_n(0)=+6$ MeV in pure neutron matter of the recent HAL-QCD
calculation [86]. Then we easily obtain that $\alpha_1/\alpha_0=+10~{\rm MeV}/-4~{\rm MeV}=-2.5$. As a result,
the ratio $U_{\Xi^-}^{\rm isovector}(r)/U_{\Xi^-}^{\rm isoscalar}(r)$ for $[(N-Z)/A]=0.2$ ($^{184}$W) is -0.5,
which is nonnegligible. But if we take the value of -24.3 MeV for the isoscalar $\Xi^-$ hyperon potential,
suggested by the very recent analysis of Ref. [99] of the $\Xi^-$ capture events in light nuclear 
emulsion identified in KEK and J-PARC experiments [42, 43], 
and retain the same value of +6 MeV as above for its potential in pure neutron
matter, we get for the ratio $U_{\Xi^-}^{\rm isovector}(r)/U_{\Xi^-}^{\rm isoscalar}(r)$ for $[(N-Z)/A]=0.2$
a substantially smaller value of -0.25.
The aforementioned means that a more realistic analysis of the $\Xi^-$ hyperon production in heavy
asymmetric targets should account for in principle not only the isoscalar $\Xi^-$-nucleus potential, but also
and an isovector one. However, since the main difference between the $\Xi^-$ nuclear potentials in light (symmetric)
and heavy (asymmetric) nuclei is coming from the Coulomb forces, which are included in our model, one may hope
that the role played by the real isovector $\Xi^-$ optical potential in this production will be moderate.
Its rigorous evaluation is beyond the scope of the present work.}$^)$
.
In addition, to extend the range of applicability of our model and to see the sensitivity of the $\Xi^-$ production cross sections from the direct processes (1), (2) to the potential $U_{\Xi^-}$, we will yet adopt in
some calculations three another additional representative options for this potential, namely: i) $U_{\Xi^-}=-25$ MeV,
ii) $U_{\Xi^-}=-15$ MeV and iii) $U_{\Xi^-}=-5$ MeV, covering in view of the above-mentioned 
the bulk of the low-energy theoretical and experimental information presently available in this field.

      The total energy $E^\prime_{h}$ of the hadron inside the nuclear matter is
expressed via its average effective mass $<m^*_{h}>$ defined above and its in-medium momentum
${\bf p}^{\prime}_{h}$ by the expression [72]:
%formula(10)
\begin{equation}
E^\prime_{h}=\sqrt{({\bf p}^{\prime}_{h})^2+(<m^*_{h}>)^2}.
\end{equation}
The momentum ${\bf p}^{\prime}_{h}$ is related to the vacuum hadron momentum ${\bf p}_{h}$
as follows [72]:
%formula(11)
\begin{equation}
E^\prime_{h}=\sqrt{({\bf p}^{\prime}_{h})^2+(<m^*_{h}>)^2}=
\sqrt{{\bf p}^2_{h}+m^2_{h}}=E_{h},
\end{equation}
where $E_{h}$ is the hadron total energy in vacuum.
%%%%%%%%%%%%%%%%%%%%%%%%%%%%%%%%%%%%%%%%%%%%%%%%%%%%%%%%%%%%%%%%%%%%%%%%%%%%%%%%%%%%%%%%%%%%%
\begin{figure}[htb]
\begin{center}
\includegraphics[width=18.0cm]{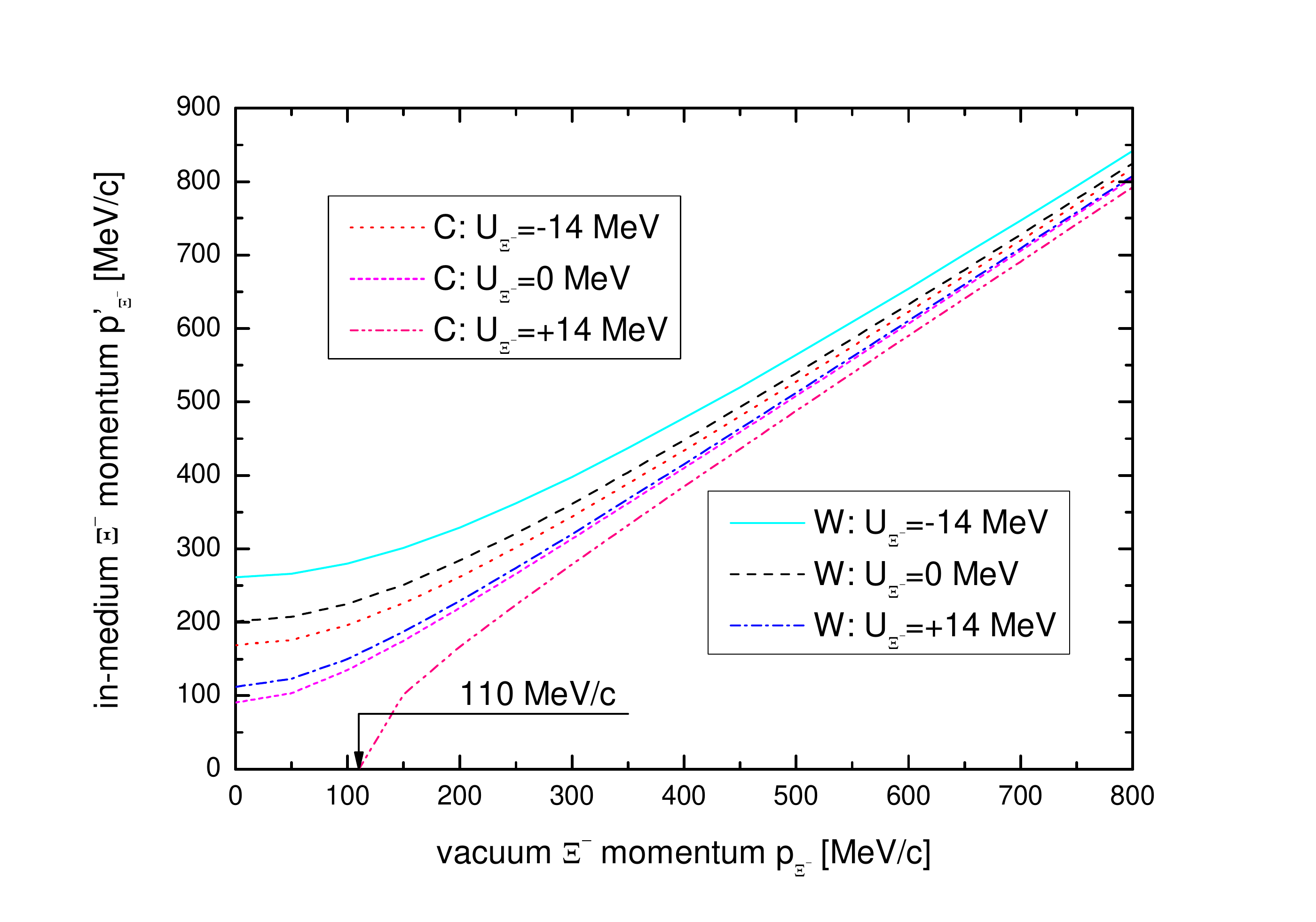}
\vspace*{-2mm} \caption{(Color online) Vacuum $\Xi^-$ hyperon momentum dependence of its in-medium
momentum for $^{12}$C and $^{184}$W nuclei at various values of the strong interaction mass shift $U_{\Xi^-}$
indicated in the inset.}
\label{void}
\end{center}
\end{figure}
%%%%%%%%%%%%%%%%%%%%%%%%%%%%%%%%%%%%%%%%%%%%%%%%%%%%%%%%%%%%%%%%%%%%%%%%%%%%%%%%%%%%%%%%%%%%%
It is of further interest to calculate, using the relations (3) and (11), the vacuum $\Xi^-$ hyperon
momentum dependence of its in-medium momentum for $^{12}$C and $^{184}$W target nuclei for the adopted
three options for the effective scalar potential $U_{\Xi^-}$. The results of such calculations are shown
in Fig. 2. We have that $p^{\prime}_{\Xi^-}(U_{\Xi^-}=-14~{\rm MeV})$ $>$ $p^{\prime}_{\Xi^-}(U_{\Xi^-}=0~{\rm MeV})$
$>$ $p^{\prime}_{\Xi^-}(U_{\Xi^-}=+14~{\rm MeV})$ $>$ $p_{\Xi^-}$ for tungsten nucleus.
In the case of carbon nucleus we also have that
$p^{\prime}_{\Xi^-}(U_{\Xi^-}=-14~{\rm MeV})$ $>$ $p^{\prime}_{\Xi^-}(U_{\Xi^-}=0~{\rm MeV})$ $>$ $p_{\Xi^-}$,
but contrary to the W nucleus case $p^{\prime}_{\Xi^-}(U_{\Xi^-}=+14~{\rm MeV})$ $<$ $p_{\Xi^-}$. The effect of the
cascade particle potential $U_{\Xi^-}$ on in-medium renormalisation of its vacuum momentum is significant at the
vacuum $\Xi^-$ momenta less than 600 MeV/c. Whereas, at higher $\Xi^-$ momenta its impact on this renormalisation
is not substantial. Thus, for example, the differences
$p^{\prime}_{\Xi^-}(U_{\Xi^-}=-14~{\rm MeV})-p^{\prime}_{\Xi^-}(U_{\Xi^-}=0~{\rm MeV})$,
$p^{\prime}_{\Xi^-}(U_{\Xi^-}=0~{\rm MeV})-p^{\prime}_{\Xi^-}(U_{\Xi^-}=+14~{\rm MeV})$ are $\sim$ 20\% and 3\%
for momenta $p_{\Xi^-}$ of 200 and 600 MeV/c, respectively, for both considered nuclei. This will lead, in particular,
to different sensitivity of the $\Xi^-$ hyperon momentum distributions to the potential $U_{\Xi^-}$, which it feels
inside the nuclear matter, at low and high $\Xi^-$ momenta (see Figs. 5 and 6 given below).

The $\Xi^-$--nucleon elastic cross section is expected to be $\sim$ 5--15 mb in the studied $\Xi^-$ momentum
range of 0.2--1.0 GeV/c [88, 89, 92, 100, 101, 102] of our main interest. The $\Xi^-$ mean "free" path up to
quasielastic rescattering can be evaluated as $\lambda^{\rm el}_{\Xi^-}=1/(<\rho_N>\sigma^{\rm el}_{\Xi^-N})$,
where $\sigma^{\rm el}_{\Xi^-N}$ is the appropriate $\Xi^-N$ elastic cross section. Using
$<\rho_N>=0.55\rho_0$ ($^{12}$C) and $<\rho_N>=0.76\rho_0$ ($^{184}$W), $\rho_0=$0.16 fm$^{-3}$ and as an
estimate of $\sigma^{\rm el}_{\Xi^-N}$ for $\Xi^-$ momenta of interest the value of 10 mb, we obtain that
$\lambda^{\rm el}_{\Xi^-}$ $\approx$ 11 fm for $^{12}$C and $\lambda^{\rm el}_{\Xi^-}$ $\approx$ 8 fm for $^{184}$W.
These values are larger, respectively, than the radii of $^{12}$C and $^{184}$W, which are approximately 3 and 7.4 fm.
Therefore, we will neglect quasielastic $\Xi^-N$ rescatterings in the present study and will do not consider
the loss and gain of the $\Xi^-$ flux in the nucleus inside the wide laboratory solid $\Xi^-$ production angle
under consideration. Then, accounting for that the in-medium threshold energies
$\sqrt{s^*_{\rm th}}=<m_{K^+}^*>+<m_{\Xi^-}^*>$ and $\sqrt{{\tilde s}^*_{\rm th}}=<m_{K^0}^*>+<m_{\Xi^-}^*>$
of the processes (1) and (2) are practically equal to each other for both considered target nuclei and the
fact that the total cross sections of these processes are described by the same functional form [60] as well as
taking into account the attenuation of the incident antikaon and the final $\Xi^-$ hyperon
in the nuclear matter in terms, respectively, of the $K^-N$ total cross section $\sigma_{{K^-}N}^{\rm tot}$
and the total inelastic $\Xi^-N$ cross section $\sigma_{{\Xi^-}N}^{\rm in}$, we represent, according to Ref. [71],
the inclusive differential cross section for the production of $\Xi^-$ hyperons with vacuum momentum
${\bf p}_{\Xi^-}$ on nuclei in the direct processes (1), (2) as follows:
%formula(12)
\begin{equation}
\frac{d\sigma_{{K^-}A \to {\Xi^-}X}^{({\rm dir})}
({\bf p}_{K^-},{\bf p}_{\Xi^-})}
{d{\bf p}_{\Xi^-}}=I_{V}[A,\theta_{\Xi^-}]
\left<\frac{d\sigma_{{K^-}p\to K^+{{\Xi^-}}}({\bf p}^{\prime}_{K^-},
{\bf p}^{\prime}_{{\Xi^-}})}{d{\bf p}^{\prime}_{{\Xi^-}}}\right>_A\frac{d{\bf p}^{\prime}_{{\Xi^-}}}
{d{\bf p}_{{\Xi^-}}},
\end{equation}
where
%formula(13)
\begin{equation}
I_{V}[A,\theta_{{\Xi^-}}]=A\int\limits_{0}^{R}r_{\bot}dr_{\bot}
\int\limits_{-\sqrt{R^2-r_{\bot}^2}}^{\sqrt{R^2-r_{\bot}^2}}dz
\rho(\sqrt{r_{\bot}^2+z^2})
\exp{\left[-\sigma_{{K^-}N}^{\rm tot}(p_{K^-})A\int\limits_{-\sqrt{R^2-r_{\bot}^2}}^{z}
\rho(\sqrt{r_{\bot}^2+x^2})dx\right]}
\end{equation}
$$
\times
\int\limits_{0}^{2\pi}d{\varphi}\exp{\left[-\sigma_{{\Xi^-}N}^{\rm in}(p^{\prime}_{\Xi^-})A
\int\limits_{0}^{l(\theta_{{\Xi^-}},\varphi)}
\rho(\sqrt{x^2+2a(\theta_{{\Xi^-}},\varphi)x+b+R^2})dx\right]};
$$
%formula(14)
\begin{equation}
a(\theta_{\Xi^-},\varphi)=z\cos{\theta_{\Xi^-}}+
r_{\bot}\sin{\theta_{\Xi^-}}\cos{\varphi},\,\,\,\,b=r_{\bot}^2+z^2-R^2,
\end{equation}
%formula(15)
\begin{equation}
l(\theta_{\Xi^-},\varphi)=\sqrt{a^2(\theta_{\Xi^-},\varphi)-b}-
a(\theta_{\Xi^-},\varphi),
\end{equation}
%formula(16)
\begin{equation}
\sigma_{{K^-}N}^{\rm tot}(p_{K^-})=
\frac{Z}{A}\sigma_{{K^-}p}^{\rm tot}(p_{K^-})+\frac{N}{A}\sigma_{{K^-}n}^{\rm tot}(p_{K^-}),
\end{equation}
%formula(17)
\begin{equation}
\sigma_{{\Xi^-}N}^{\rm in}(p^{\prime}_{\Xi^-})=
\frac{Z}{A}\sigma_{{\Xi^-}p}^{\rm in}(p^{\prime}_{\Xi^-})+
\frac{N}{A}\sigma_{{\Xi^-}n}^{\rm in}(p^{\prime}_{\Xi^-})
\end{equation}
and
%formula(18)
\begin{equation}
\left<\frac{d\sigma_{{K^-}p\to K^+{\Xi^-}}({\bf p}^{\prime}_{K^-},{\bf p}^{\prime}_{\Xi^-})}
{d{\bf p}^{\prime}_{\Xi^-}}\right>_A=
\int\int
P_A({\bf p}_t,E)d{\bf p}_tdE
\end{equation}
$$
\times
\left\{\frac{d\sigma_{{K^-}p\to K^+{\Xi^-}}[\sqrt{s^*},<m_{K^+}^*>,
<m_{\Xi^-}^*>,{\bf p}^{\prime}_{\Xi^-}]}
{d{\bf p}^{\prime}_{\Xi^-}}\right\},
$$
%formula(19)
\begin{equation}
  s^*=(E^{\prime}_{K^-}+E_t)^2-({\bf p}^{\prime}_{K^-}+{\bf p}_t)^2,
\end{equation}
%formula(20)
\begin{equation}
   E_t=M_A-\sqrt{(-{\bf p}_t)^2+(M_{A}-m_{N}+E)^{2}}.
\end{equation}
Here,
$d\sigma_{{K^-}p\to {K^+}{\Xi^-}}[\sqrt{s^*},<m_{K^+}^*>,<m_{\Xi^-}^*>,{\bf p}^{\prime}_{\Xi^-}]
/d{\bf p}^{\prime}_{\Xi^-}$
is the off-shell inclusive differential cross section for the production of $K^+$ meson and
$\Xi^-$ hyperon with modified masses $<m_{K^+}^*>$ and $<m_{\Xi^-}^*>$, respectively.
The $\Xi^-$ hyperon is produced with in-medium momentum
${\bf p}^{\prime}_{{\Xi^-}}$ in process (1) at the ${K^-}p$ center-of-mass energy $\sqrt{s^*}$.
$E^{\prime}_{K^-}$ and ${\bf p}^{\prime}_{K^-}$ are in-medium total energy and momentum of the incident antikaon,
which are related by the equation (10);
$\rho({\bf r})$ and $P_A({\bf p}_t,E)$ are the local nucleon density and the
spectral function of the target nucleus A normalized to unity
(the detailed information about these quantities, used in our calculations, is given
in Refs. [71, 75, 103, 104]);
${\bf p}_t$ and $E$ are the internal momentum and removal energy of the target proton
involved in the collision process (1);
$Z$ and $N$ are the numbers of protons and neutrons in the target nucleus ($A=Z+N$),
$M_{A}$  and $R$ are its mass and radius; $m_N$ is the free space nucleon mass;
$\theta_{\Xi^-}$ is the polar angle of
vacuum momentum ${\bf p}_{{\Xi^-}}$ in the laboratory system with z-axis directed along the vacuum
momentum ${\bf p}_{{K^-}}$ of the incident antikaon beam;
$\sigma_{{K^-}p}^{\rm tot}(p_{K^-})$ and $\sigma_{{K^-}n}^{\rm tot}(p_{K^-})$
are the total cross section of the free [105, 106] ${K^-}p$ and ${K^-}n$ interactions
\footnote{$^)$ It should be mentioned that the use of the total ${K^-}p$ and ${K^-}n$
cross sections instead of the inelastic ones in Eqs. (13), (16)
is caused by the fact that the $K^-$ energy loss in nucleus, associated with the $K^-$ meson
quasielastic rescatterings on the intranuclear nucleons, will lead to a substantial reduction
of its possibility to create a $\Xi^-$ hyperon in the processes (1) and (2) at considered near-threshold
laboratory incident antikaon momenta, or in other words, will cause the additional to the genuine
"absorption" of the $K^-$ meson flux in nuclear matter with respect to the $\Xi^-$ creation in
these processes.}$^)$
at vacuum beam momenta of 1.0 and 1.3 GeV/c
\footnote{$^)$ We assume that $\sigma_{{K^-}p}^{\rm tot}(p^{\prime}_{K^-})$ $\approx$
$\sigma_{{K^-}p}^{\rm tot}(p_{K^-})$ and $\sigma_{{K^-}n}^{\rm tot}(p^{\prime}_{K^-})$ $\approx$
$\sigma_{{K^-}n}^{\rm tot}(p_{K^-})$. This is well justified due to the following.
For example, for incident $K^-$ meson momentum of $p_{K^-}$=1.0 GeV/c
its in-medium momentum $p^{\prime}_{K^-}$ is equal to 1.012 and 1.021 GeV/c for $^{12}$C and $^{184}$W,
respectively. And according to [107], we have that
$\sigma_{{K^-}p(K^-n)}^{\rm tot}(p^{\prime}_{K^-}=1.012, 1.021~{\rm GeV/c})$ $\approx$
$\sigma_{{K^-}p(K^-n)}^{\rm tot}(p_{K^-}=1.0~{\rm GeV/c})$.}$^)$
.
At these momenta, $\sigma_{{K^-}p}^{\rm tot}\approx$ 50 and 30 mb. And
$\sigma_{{K^-}n}^{\rm tot}\approx$ 40 and 30 mb [107]. With these, the quantity $\sigma_{{K^-}N}^{\rm tot}$,
entering into Eqs. (13) and (16), amounts approximately to 45 and 30 mb for initial momenta of
1.0 and 1.3 GeV/c for both considered target nuclei $^{12}$C$_6$ and $^{184}$W$_{74}$.
We will employ these values in our calculations. In Eq. (12) it is assumed that the $\Xi^-$ hyperon
production cross sections in $K^-p$ and $K^-n$ interactions (1) and (2) are the same [60].
In addition, it is suggested that the ways of the incident antikaon from the vacuum to the production
point inside the nucleus and of the produced cascade hyperon at this point out of the nucleus are not
disturbed by the weak $K^-$ and $\Xi^-$ nuclear potentials, considered in the present work, by the respective
attractive Coulomb potentials as well as by rare $K^-N$ and ${\Xi^-}N$ elastic rescatterings. As a result,
the in-medium antikaon and hyperon momenta ${\bf p}^{\prime}_{K^-}$ and ${\bf p}^{\prime}_{\Xi^-}$
are assumed to be parallel, respectively, to the vacuum ones ${\bf p}_{K^-}$ and ${\bf p}_{\Xi^-}$
and the relation between them is given by Eq. (11).

   In line with [71, 72], we suppose that the off-shell differential cross section
$d\sigma_{{K^-}p \to {K^+}{\Xi^-}}[\sqrt{s^*},<m_{K^+}^*>,<m_{\Xi^-}^*>,{\bf p}^{\prime}_{\Xi^-}]
/d{\bf p}^{\prime}_{\Xi^-}$ for $\Xi^-$ production in process (1) is equivalent to
the respective on-shell cross section calculated for the off-shell kinematics of this process
as well as for the final $K^+$ meson and $\Xi^-$ hyperon in-medium masses $<m_{K^+}^*>$
and $<m_{\Xi^-}^*>$, respectively. Accounting for the results given in Ref. [71] and assuming that the
$\Xi^-$ angular distribution in reaction (1) is isotropic in the ${K^-}p$ c.m.s. at beam momenta
of interest (cf. [98, 108]
\footnote{$^)$ Presented here experimental angular distribution of the reaction $K^-p \to K^+\Xi^-$
shows practically an isotropic behavior at the lowest c.m.s. collision energy of 1.95 GeV (or at the
lowest $K^-$ beam momentum of 1.34 GeV/c).}$^)$
),
we obtain the following expression for this differential cross section:
%FORMULA (21)
\begin{equation}
\frac{d\sigma_{K^{-}p \to {K^+}{\Xi^-}}[\sqrt{s^*},<m_{K^+}^*>,<m_{\Xi^-}^*>,{\bf p}^{\prime}_{\Xi^-}]}
{d{\bf p}^{\prime}_{\Xi^-}}=
\frac{\pi}{I_2[s^*,<m_{K^+}^*>,<m_{\Xi^-}^*>]E^{\prime}_{\Xi^-}}
\end{equation}
$$
\times
\frac{\sigma_{{K^{-}}p \to {K^+}{\Xi^-}}(\sqrt{s^*},\sqrt{s^*_{\rm th}})}{4\pi}
$$
$$
\times
\frac{1}{(\omega+E_t)}\delta\left[\omega+E_t-\sqrt{(<m_{K^+}^*>)^2+({\bf Q}+{\bf p}_t)^2}\right],
$$
where
%FORMULA (22)
\begin{equation}
I_2[s^*,<m_{K^+}^*>,<m_{\Xi^-}^*>]=\frac{\pi}{2}
\frac{\lambda[s^*,(<m_{K^+}^*>)^{2},(<m_{\Xi^-}^*>)^{2}]}{s^*},
\end{equation}
%FORMULA (23)
\begin{equation}
\lambda(x,y,z)=\sqrt{{\left[x-({\sqrt{y}}+{\sqrt{z}})^2\right]}{\left[x-
({\sqrt{y}}-{\sqrt{z}})^2\right]}},
\end{equation}
%FORMULA (24)
\begin{equation}
\omega=E^{\prime}_{K^-}-E^{\prime}_{\Xi^-}, \,\,\,\,{\bf Q}={\bf p}^{\prime}_{K^-}-{\bf p}^{\prime}_{\Xi^-}.
\end{equation}
Here, $\sigma_{{K^{-}}p \to {K^+}{\Xi^-}}(\sqrt{s^*},\sqrt{s^*_{\rm th}})$ is the
"in-medium" total cross section of reaction (1) having the threshold energy $\sqrt{s^*_{\rm th}}$ defined above.
In line with the above-mentioned, it is equivalent to the vacuum cross section
$\sigma_{{K^{-}}p \to {K^+}{\Xi^-}}(\sqrt{s},\sqrt{s_{\rm th}})$, in which the vacuum threshold energy
$\sqrt{s_{\rm th}}=m_{K^+}+m_{\Xi^-}=1.8154$ GeV is replaced by the in-medium one $\sqrt{s^*_{\rm th}}$
and the free center-of-mass energy squared $s$, presented by the formula
%FORMULA (25)
\begin{equation}
s=(E_{K^-}+m_N)^2-{\bf p}_{K^-}^2,
\end{equation}
is replaced by the in-medium one $s^*$, defined by the expression (19).

For the free total cross section
$\sigma_{{K^{-}}p \to {K^+}{\Xi^-}}(\sqrt{s},\sqrt{s_{\rm th}})$ we have adopted the
following parametrization, suggested in Ref. [60]:
%formula(26)
\begin{equation}
\sigma_{{K}^-p \to {K^+}{\Xi^-}}(\sqrt{s},\sqrt{s_{\rm th}})=
	235.6\left(1-\sqrt{s_{\rm th}}/\sqrt{s}\right)^{2.4}
               \left(\sqrt{s_{\rm th}}/\sqrt{s}\right)^{16.6}~[{\rm mb}].
\end{equation}
\begin{figure}[htb]
\begin{center}
\includegraphics[width=18.0cm]{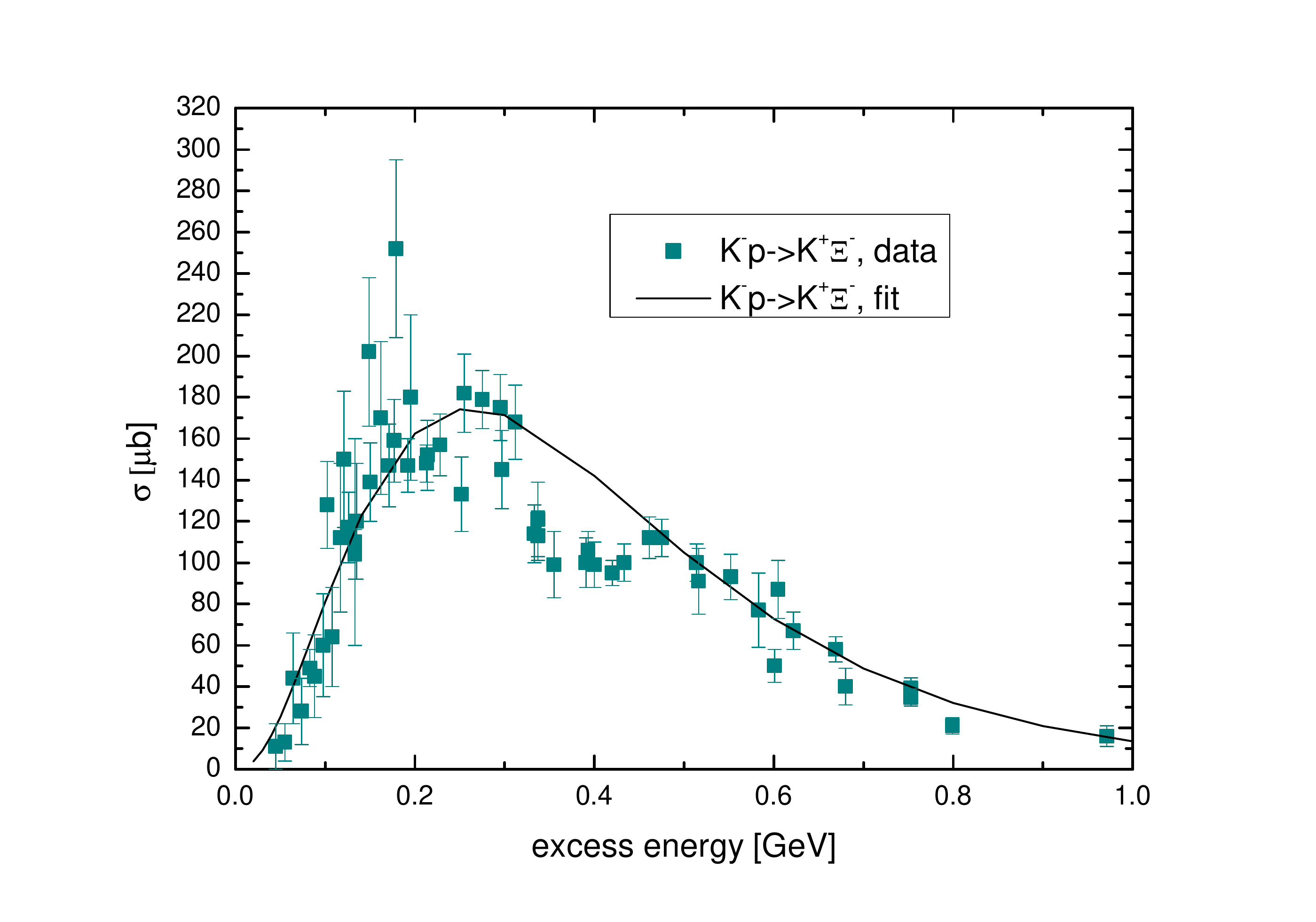}
\vspace*{-2mm} \caption{(Color online) Total cross section for the reaction $K^-p \to {K^+}{\Xi^-}$
as function of the available excess energy $\sqrt{s}-\sqrt{s_{\rm th}}$
above the threshold $\sqrt{s_{\rm th}}$. For notation see the text.}
\label{void}
\end{center}
\end{figure}
%%%%%%%%%%%%%%%%%%%%%%%%%%%%%%%%%%%%%%%%%%%%%%%%%%%%%%%%%%%%%%%%%%%%%%%%%%%%%%%%%%%%%%%%%%%%%
As is seen from Fig. 3, it (solid curve) fits well the data (full squares),
taken from the compilation of Flaminio {\it et al.} [107].
As can also be seen from this figure, the on-shell cross section $\sigma_{{K}^-p \to {K^+}{\Xi^-}}$
amounts approximately to 100 ${\rm \mu}$b for the initial antikaon momentum of 1.3 GeV/c, corresponding
to the excess energy $\sqrt{s}-\sqrt{s_{\rm th}}$=117 MeV.
This offers the possibility of measuring the $\Xi^-$ yield in near-threshold $K^-A$ reactions
at the J-PARC Hadron Experimental Facility with sizable strength using high-intensity separated secondary
antikaon beams in the K1.8 and K1.8BR beamlines [48].
%%%%%%%%%%%%%%%%%%%%%%%%%%%%%%%%%%%%%%%%%%%%%%%%%%%%%%%%%%%%%%%%%%%%%%%%%%%%%%%%%%%%%%%%%%%%%
\begin{figure}[htb]
\begin{center}
\includegraphics[width=18.0cm]{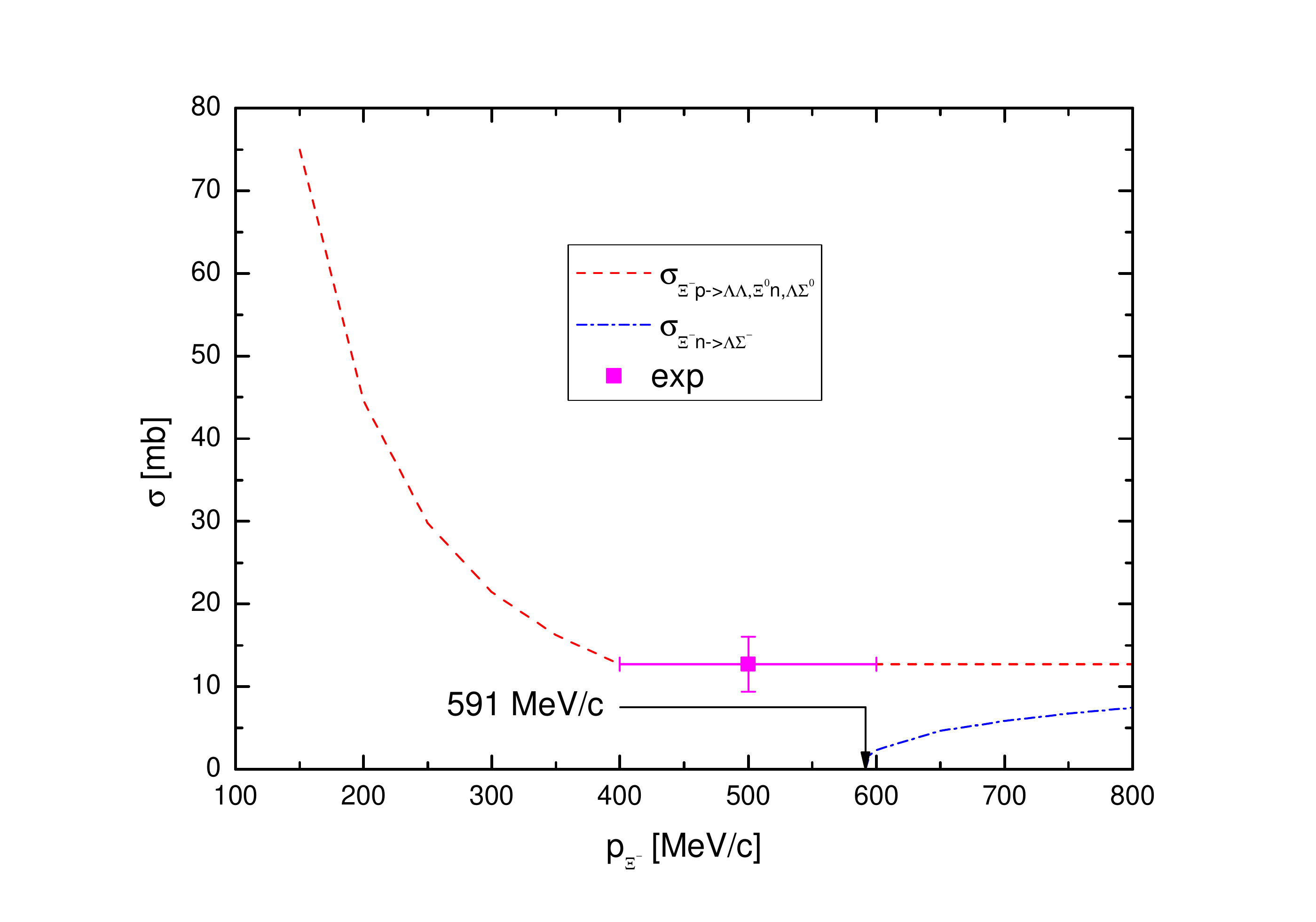}
\vspace*{-2mm} \caption{(Color online) $\Xi^-$ hyperon momentum dependence of the total
cross sections for the reactions $\Xi^-p \to {\Lambda}{\Lambda},{\Xi^0}{n},{\Lambda}{\Sigma^0}$
and $\Xi^-n \to {\Lambda}{\Sigma^-}$. The arrow indicates the free threshold momentum for the latter one.
For the rest of the notation see the text.}
\label{void}
\end{center}
\end{figure}
%%%%%%%%%%%%%%%%%%%%%%%%%%%%%%%%%%%%%%%%%%%%%%%%%%%%%%%%%%%%%%%%%%%%%%%%%%%%%%%%%%%%%%%%%%%%%
%%%%%%%%%%%%%%%%%%%%%%%%%%%%%%%%%%%%%%%%%%%%%%%%%%%%%%%%%%%%%%%%%%%%%%%%%%%%%%%%%%%%%%%%%%%%%
\begin{figure}[!h]
\begin{center}
\includegraphics[width=18.0cm]{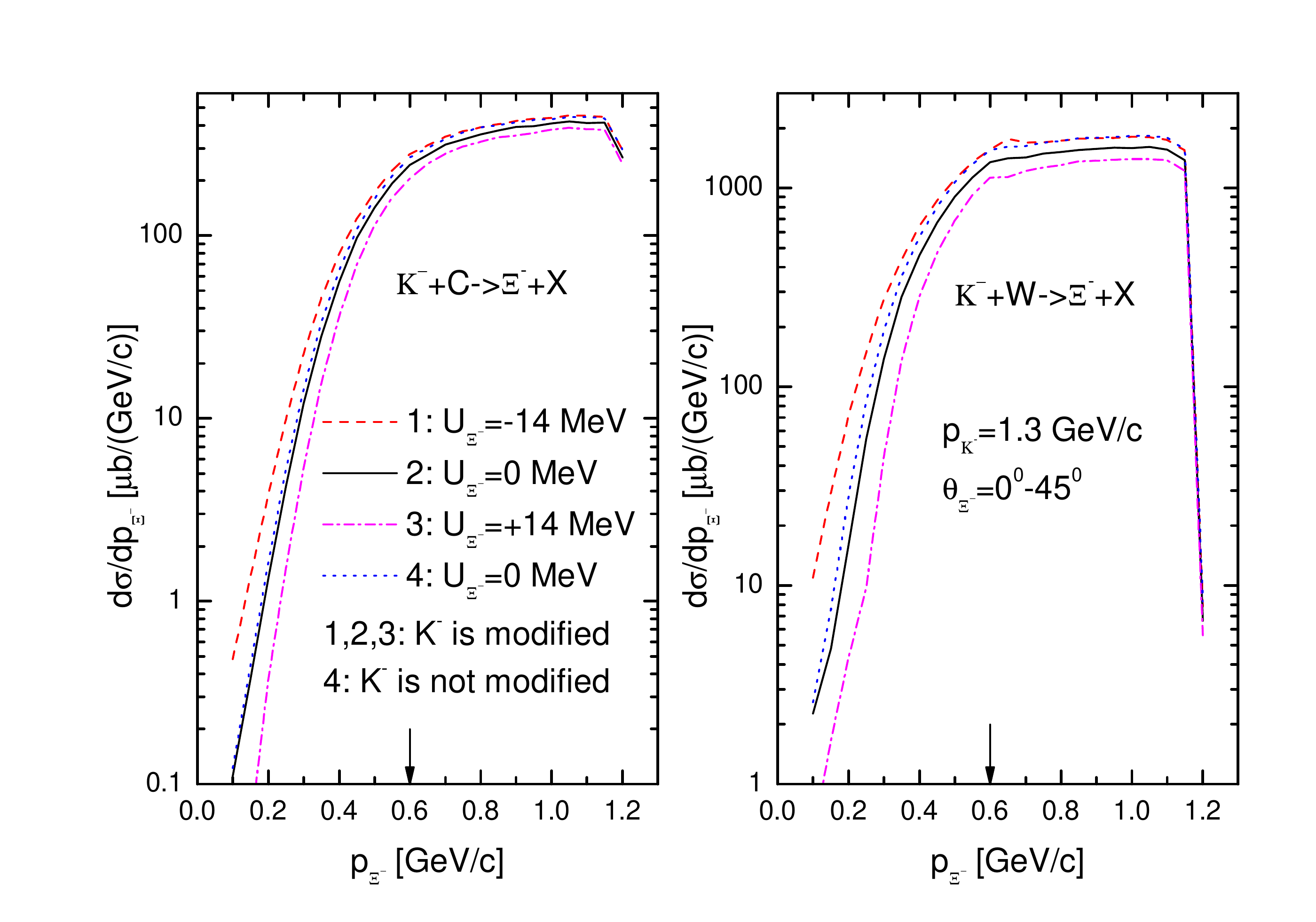}
\vspace*{-2mm} \caption{(Color online) Momentum differential cross sections for the production of $\Xi^-$
hyperons from the direct ${K^-}p \to {K^+}{\Xi^-}$ and ${K^-}n \to {K^0}{\Xi^-}$
processes in the laboratory polar angular range of
0$^{\circ}$--45$^{\circ}$ in the interaction of medium-modified and not modified
$K^-$ mesons having vacuum momentum of 1.3 GeV/c with $^{12}$C
(left) and $^{184}$W (right) nuclei, calculated for different values of the $\Xi^-$ hyperon effective
scalar potential $U_{\Xi^-}$ at density $\rho_0$ indicated in the inset and
for the nominal $\Xi^-$ absorption in the nuclear matter.
The arrows indicate the boundary between the low-momentum and high-momentum parts of the $\Xi^-$ spectra.}
\label{void}
\end{center}
\end{figure}
%%%%%%%%%%%%%%%%%%%%%%%%%%%%%%%%%%%%%%%%%%%%%%%%%%%%%%%%%%%
%\begin{figure}[htb]
%\begin{center}
%\includegraphics[width=18.0cm]{fig3kamincas.pdf}
%\vspace*{-2mm} \caption{(Color online) $\Xi^-$ hyperon in-medium momentum dependence of the total
%cross sections for the reactions $\Xi^-p \to {\Lambda}{\Lambda},{\Xi^0}{n},{\Lambda}{\Sigma^0}$
%and $\Xi^-n \to {\Lambda}{\Sigma^-}$ (cf. Fig. 2) as well as of the total cross sections for these
%reactions, averaged over target nucleon binding and Fermi motion.}
%\label{void}
%\end{center}
%\end{figure}
%%%%%%%%%%%%%%%%%%%%%%%%%%%%%%%%%%%%%%%%%%%%%%%%%%%%%%%%%%%%%%%%%%%%%%%%%%%%%%%%%%%%%%%%%%%%%

Accounting for the fact that in the considered initial momentum
region $\Xi^-$ hyperons are mainly emitted, due to the kinematics, in a narrow cone along the beam line
\footnote{$^)$ Thus, for instance, at a beam momentum of 1.3 GeV/c the $\Xi^-$ laboratory production
polar angles in reaction (1) proceeding on the target proton being at rest are $\le$ 19.65$^{\circ}$.}$^)$
,
we will calculate the $\Xi^-$ momentum
differential and total production cross sections on $^{12}$C and $^{184}$W targets
for the laboratory solid angle
${\Delta}{\bf \Omega}_{\Xi^-}$=$0^{\circ} \le \theta_{\Xi^-} \le 45^{\circ}$,
and $0 \le \varphi_{\Xi^-} \le 2{\pi}$.
Here, $\varphi_{\Xi^-}$ is the azimuthal angle of the $\Xi^-$ momentum ${\bf p}_{\Xi^-}$
in the laboratory system.
Then, integrating the full inclusive differential cross section (12) over this angular domain with
taking into account that in Eqs. (12)--(15) we suppose that the direction of the $\Xi^-$ hyperon momentum
is not changed during its propagation inside the nucleus from the production point in it to the vacuum,
we can represent the differential cross section for $\Xi^-$ hyperon
production in antikaon-induced reactions from the direct processes (1), (2),
corresponding to this angle, in the following form:
%formula(27)
\begin{equation}
\frac{d\sigma_{{K^-}A\to {\Xi^-}X}^{({\rm dir})}
(p_{K^-},p_{\Xi^-})}{dp_{\Xi^-}}=
\int\limits_{{\Delta}{\bf \Omega}_{\Xi^-}}^{}d{\bf \Omega}_{\Xi^-}
\frac{d\sigma_{{K^-}A\to {\Xi^-}X}^{({\rm dir})}
({\bf p}_{K^-},{\bf p}_{\Xi^-})}{d{\bf p}_{\Xi^-}}p_{\Xi^-}^2
\end{equation}
$$
=2{\pi}\left(\frac{p_{\Xi^-}}{p^{\prime}_{\Xi^-}}\right)
\int\limits_{\cos45^{\circ}}^{1}d\cos{{\theta_{\Xi^-}}}I_{V}[A,\theta_{\Xi^-}]
\left<\frac{d\sigma_{{K^-}p\to {K^+}{\Xi^-}}(p^{\prime}_{K^-},
p^{\prime}_{\Xi^-},\theta_{\Xi^-})}{dp^{\prime}_{\Xi^-}d{\bf \Omega}_{\Xi^-}}\right>_A.
$$

We will consider also the effects from $\Xi^-$ in-medium modification and from $\Xi^-$ absorption in nuclear
matter on the momentum dependence of the following relative observable -- the transparency ratio $T_A$
defined as the ratio between the inclusive differential cascade production cross section (27) on a heavy
nucleus and a light one ($^{12}$C):
%formula(28)
\begin{equation}
T_A(p_{K^-},p_{\Xi^-})=\frac{12}{A}\frac{d\sigma_{{K^-}A\to {\Xi^-}X}^{({\rm dir})}
(p_{K^-},p_{\Xi^-})/dp_{\Xi^-}}
{d\sigma_{{K^-}{\rm C}\to {\Xi^-}X}^{({\rm dir})}
(p_{K^-},p_{\Xi^-})/dp_{\Xi^-}}.
\end{equation}

We define now the total inelastic $\Xi^-p$ and $\Xi^-n$ cross sections
$\sigma_{{\Xi^-}p}^{\rm in}$ and $\sigma_{{\Xi^-}n}^{\rm in}$,
appearing in Eq. (17) and used in our calculations of $\Xi^-$ production in $K^-A$ reactions.
At $\Xi^-$ momenta below 1 GeV/c of interest these cross sections are entirely exhausted by the
total cross sections $\sigma_{\Xi^-p \to {\Lambda}\Lambda}$, $\sigma_{\Xi^-p \to {\Xi^0}n}$,
$\sigma_{\Xi^-p \to {\Lambda}\Sigma^0}$ and $\sigma_{\Xi^-n \to {\Lambda}\Sigma^-}$ of the
processes $\Xi^-p \to {\Lambda}\Lambda$, $\Xi^-p \to {\Xi^0}n$, $\Xi^-p \to {\Lambda}\Sigma^0$ and
$\Xi^-n \to {\Lambda}\Sigma^-$. While the first two channels are open at any finite $\Xi^-$ momentum
in free scattering, the last two inelastic ones open only at momenta of about 0.57 and 0.59 GeV/c,
respectively, in this scattering. At considered $\Xi^-$ momenta we can ignore the processes
$\Xi^-p \to {\Lambda}\Lambda{\pi^0}$, $\Xi^-N \to {\Lambda}\Sigma{\pi}$ and $\Xi^-N \to {\Sigma}\Sigma$,
since they only open at threshold momenta around 1 GeV/c. Thus, these momenta in free $\Xi^-N$ interactions
are about 0.87, 1.20 and 0.96 GeV/c, respectively. With these, we have
%formula(29)
\begin{equation}
\sigma_{{\Xi^-}p}^{\rm in}=\sigma_{\Xi^-p \to {\Lambda}\Lambda,{\Xi^0}n,\Lambda{\Sigma^0}}=
\sigma_{\Xi^-p \to {\Lambda}\Lambda}+\sigma_{\Xi^-p \to {\Xi^0}n}+\sigma_{\Xi^-p \to {\Lambda}\Sigma^0},
\end{equation}
%formula(30)
\begin{equation}
\sigma_{{\Xi^-}n}^{\rm in}=\sigma_{\Xi^-n \to {\Lambda}\Sigma^-}.
\end{equation}
The cross section $\sigma_{\Xi^-p \to {\Lambda}\Lambda,{\Xi^0}n,\Lambda{\Sigma^0}}$ was calculated in
[88] at laboratory $\Xi^-$ momenta $\le$ 800 MeV/c within the updated version chiral EFT at the NLO for
${\Xi}N$ interaction, which is in line with the empirical information on the ${\Lambda}\Lambda$ $S$-wave
scattering length [89] and fulfills all available scarce experimental constraints [100, 101, 109, 110, 111]
on the ${\Xi^-}p$ elastic and inelastic
cross sections and leads to a weakly attractive single-particle potential of the $\Xi^-$ hyperon in
nuclear matter as evidenced by recent observation of the existence of $\Xi^-$ hypernuclei (see above).
We parametrize the results of calculations [88] by the following laboratory $\Xi^-$ momentum $p_{\Xi^-}$
dependence:
%formula(31)
\begin{equation}
\sigma_{{\Xi^-}p}^{\rm in}(p_{\Xi^-})=
\sigma_{\Xi^-p \to {\Lambda}\Lambda,{\Xi^0}n,\Lambda{\Sigma^0}}(p_{\Xi^-})=\left\{
\begin{array}{ll}
	2.417/(p_{\Xi^-})^{1.8106}~[{\rm mb}]
	&\mbox{for $0.15 \le p_{\Xi^-} \le 0.4~{\rm GeV/c}$}, \\
	&\\
                   12.7~[{\rm mb}]
	&\mbox{for $0.4 < p_{\Xi^-} \le 0.8~{\rm GeV/c}$},
\end{array}
\right.	
\end{equation}
where momentum $p_{\Xi^-}$ is expressed in GeV/c. For illustration this dependence
is shown in Fig. 4 by the dashed curve. As visible in Fig. 4, it fits well the available data point
taken from [110] over the momentum range $400 < p_{\Xi^-} < 600~{\rm MeV/c}$. The on-shell cross section
of $\Xi^0p \to {\Lambda}\Sigma^+$ (or due to the isospin symmetry of $\Xi^-n \to {\Lambda}\Sigma^-$)
reaction, calculated within the previous version chiral EFT at the NLO [89], was parametrized by us
as follows:
%formula(32)
\begin{equation}
\sigma_{{\Xi^-}n}^{\rm in}(p_{\Xi^-})=\sigma_{\Xi^-n \to {\Lambda}\Sigma^-}(p_{\Xi^-})=
23.448\left(\sqrt{s_{\Xi^-}(p_{\Xi^-})}-\sqrt{s_0}\right)^{0.353}~[{\rm mb}],
\end{equation}
where
%FORMULA (33)
\begin{equation}
s_{\Xi^-}(p_{\Xi^-})=(E_{\Xi^-}+m_N)^2-p_{\Xi^-}^2
\end{equation}
and $\sqrt{s_0}$=$m_{\Lambda}$+$m_{\Sigma^-}$=2.313132 GeV is the free threshold energy for the
$\Xi^-n \to {\Lambda}\Sigma^-$ reaction. For completeness, this parametrization is shown in Fig. 4
by the dotted-dashed curve. For the in-medium ${\Xi^-}p$ and ${\Xi^-}n$ inelastic cross sections
we use Eqs. (31) and (32), in which one needs to make only the substitution
$p_{\Xi^-}$ $\to$ $p^{\prime}_{\Xi^-}$.
Following the predictions of the approaches [85, 86, 87, 112, 113] that the $\Lambda$ and $\Sigma$
hyperons experience only a moderately attractive and repulsive potentials $\sim$
-(10--20) MeV and +(10--20) MeV, respectively, at central densities and relevant $\Lambda$ and
$\Sigma$ finite momenta $\sim$, as showed our estimates, 300 MeV/c, we assume that the threshold energy
$\sqrt{s_0}$ entering into Eq. (32) is not changed in the nuclear medium due to cancellation of these
potentials. To study the sensitivity of the $\Xi^-$ hyperon production cross sections from processes (1)
and (2) to its absorption in nuclear matter we will also employ in our calculations yet two additional
scenarios for the $\Xi^-p$ and $\Xi^-n$ nominal inelastic cross sections (31) and (32), in which these
cross sections were artificially multiplied by factors $f=0.5$ and $f=2$.
%%%%%%%%%%%%%%%%%%%%%%%%%%%%%%%%%%%%%%%%%%%%%%%%%%%%%%%%%%%
\begin{figure}[!h]
\begin{center}
\includegraphics[width=18.0cm]{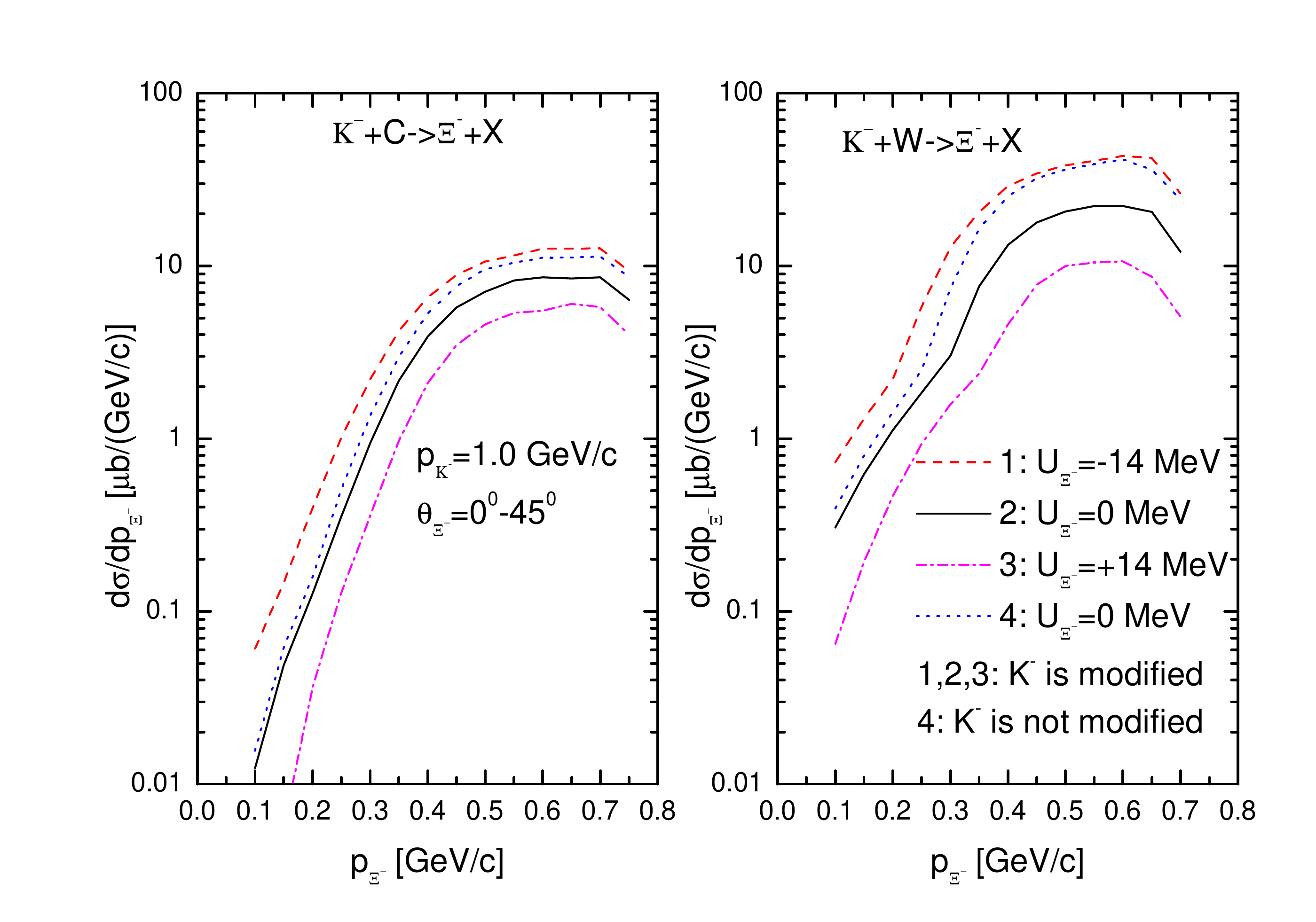}
\vspace*{-2mm} \caption{(Color online) The same as in Fig. 5,
but for the initial vacuum antikaon momentum of 1.0 GeV/c.}
\label{void}
\end{center}
\end{figure}
%%%%%%%%%%%%%%%%%%%%%%%%%%%%%%%%%%%%%%%%%%%%%%%%%%%%%%%%%%%
%%%%%%%%%%%%%%%%%%%%%%%%%%%%%%%%%%%%%%%%%%%%%%%%%%%%%%%%%%%
\begin{figure}[htb]
\begin{center}
\includegraphics[width=18.0cm]{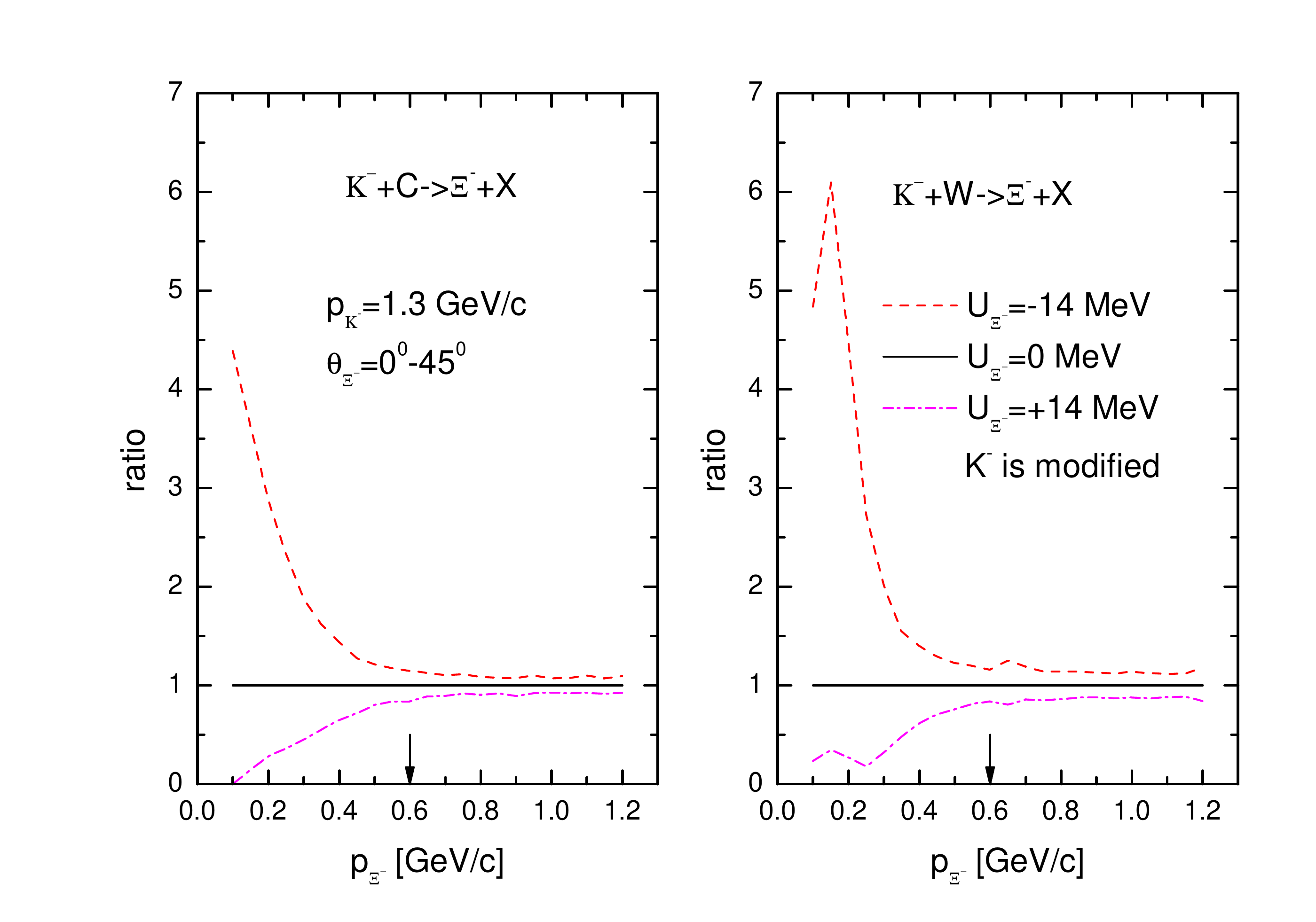}
\vspace*{-2mm} \caption{(Color online) Ratio between the differential cross sections for $\Xi^-$
production on $^{12}$C (left) and $^{184}$W (right) target nuclei in the angular region of
0$^{\circ}$--45$^{\circ}$ by medium-modified $K^-$ mesons having vacuum momentum of 1.3 GeV/c,
calculated with and without the $\Xi^-$ hyperon in-medium mass shift
$U_{\Xi^-}$ at density $\rho_0$ indicated in the inset and for the nominal $\Xi^-$ absorption
in the nuclear matter, as a function of $\Xi^-$ momentum.
The arrows indicate the boundary between the low-momentum and high-momentum parts of the
$\Xi^-$ spectra.}
\label{void}
\end{center}
\end{figure}
%%%%%%%%%%%%%%%%%%%%%%%%%%%%%%%%%%%%%%%%%%%%%%%%%%%%%%%%%%%
\begin{figure}[!h]
\begin{center}
\includegraphics[width=18.0cm]{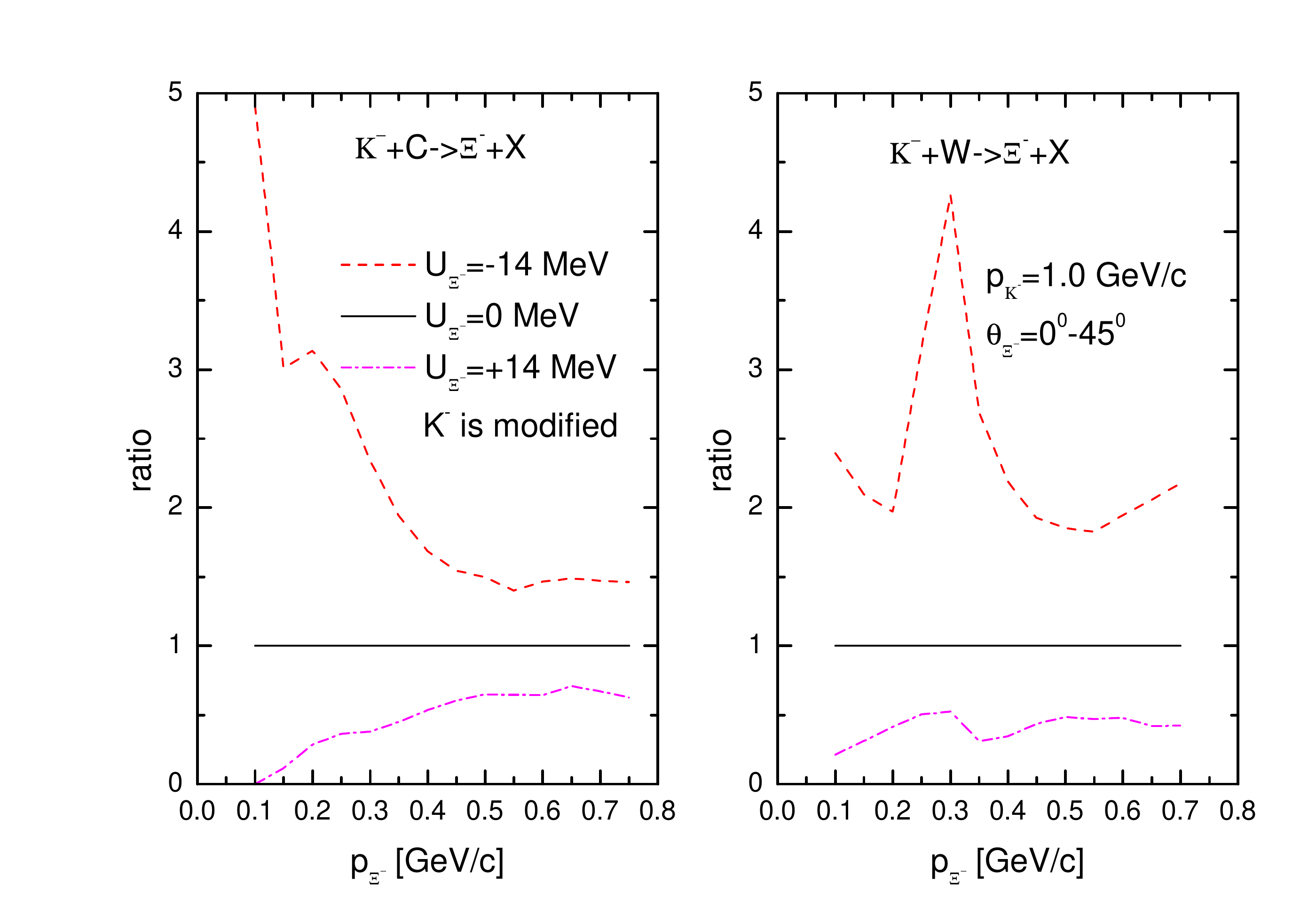}
\vspace*{-2mm} \caption{(Color online) The same as in Fig. 7,
but for the initial vacuum antikaon momentum of 1.0 GeV/c.}
\label{void}
\end{center}
\end{figure}
%%%%%%%%%%%%%%%%%%%%%%%%%%%%%%%%%%%%%%%%%%%%%%%%%%%%%%%%%%%
%%%%%%%%%%%%%%%%%%%%%%%%%%%%%%%%%%%%%%%%%%%%%%%%%%%%%%%%%%%%%%%%%%%%%%%%%%%%%%%%%%%%%%%%%%%%%
\begin{figure}[!h]
\begin{center}
\includegraphics[width=18.0cm]{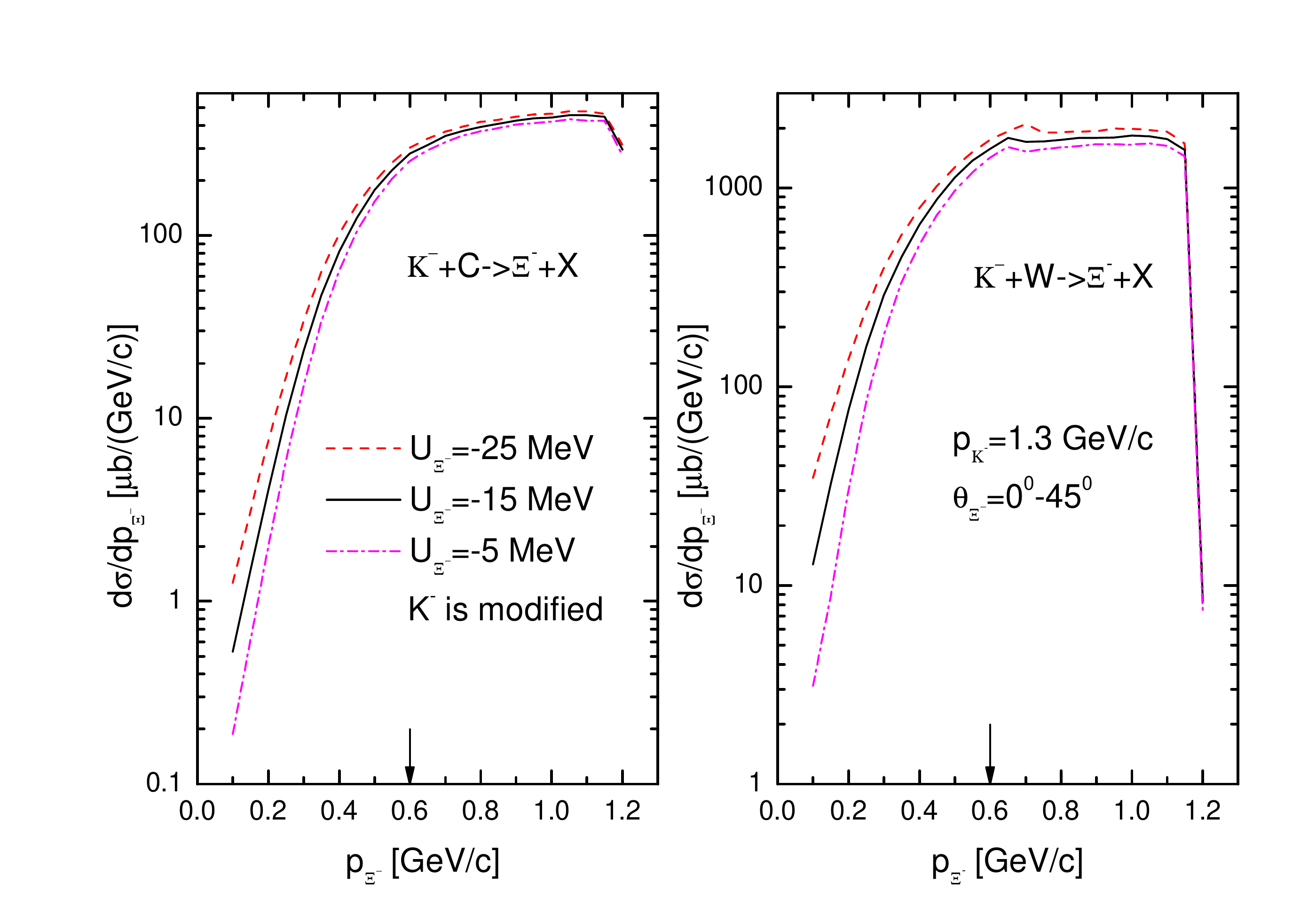}
\vspace*{-2mm} \caption{(Color online) Momentum differential cross sections for the production of $\Xi^-$
hyperons from the direct ${K^-}p \to {K^+}{\Xi^-}$ and ${K^-}n \to {K^0}{\Xi^-}$
processes in the laboratory polar angular range of
0$^{\circ}$--45$^{\circ}$ in the interaction of medium-modified
$K^-$ mesons having vacuum momentum of 1.3 GeV/c with $^{12}$C
(left) and $^{184}$W (right) nuclei, calculated for another compared to the above
different values of the $\Xi^-$ hyperon effective
scalar potential $U_{\Xi^-}$ at density $\rho_0$ indicated in the inset and
for the nominal $\Xi^-$ absorption in the nuclear matter.
The arrows indicate the boundary between the low-momentum and high-momentum parts of the $\Xi^-$ spectra.}
\label{void}
\end{center}
\end{figure}
%%%%%%%%%%%%%%%%%%%%%%%%%%%%%%%%%%%%%%%%%%%%%%%%%%%%%%%%%%%
%%%%%%%%%%%%%%%%%%%%%%%%%%%%%%%%%%%%%%%%%%%%%%%%%%%%%%%%%%%
\begin{figure}[htb]
\begin{center}
\includegraphics[width=18.0cm]{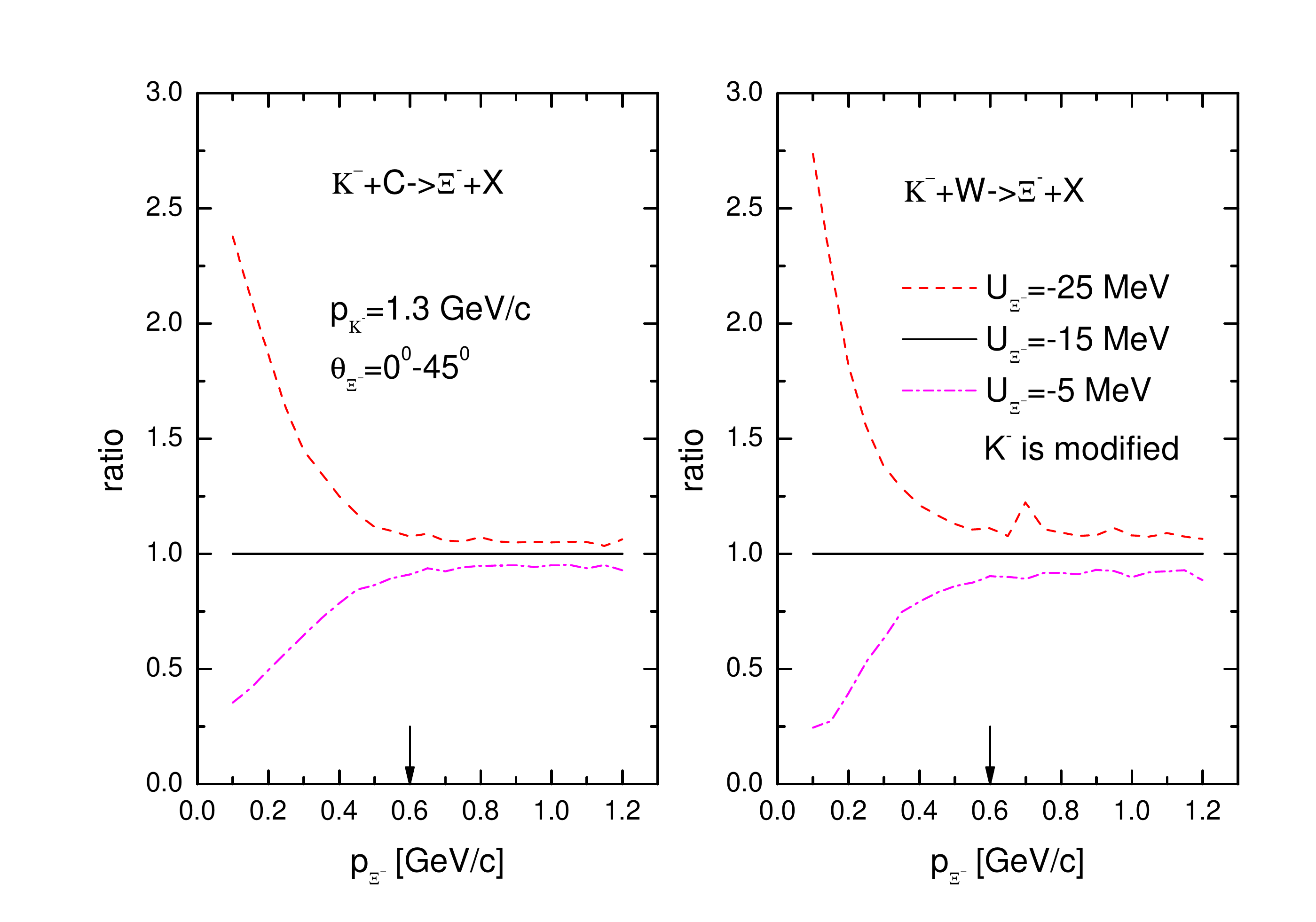}
\vspace*{-2mm} \caption{(Color online) Ratio between the differential cross sections for $\Xi^-$
production on $^{12}$C (left) and $^{184}$W (right) target nuclei in the angular region of
0$^{\circ}$--45$^{\circ}$ by medium-modified $K^-$ mesons having vacuum momentum of 1.3 GeV/c,
calculated with the $\Xi^-$ hyperon in-medium mass shift
$U_{\Xi^-}$ at density $\rho_0$ indicated in the inset and with the shift $U_{\Xi^-}=-15$ MeV
for the nominal $\Xi^-$ absorption
in the nuclear matter, as a function of $\Xi^-$ momentum.
The arrows indicate the boundary between the low-momentum and high-momentum parts of the
$\Xi^-$ spectra.}
\label{void}
\end{center}
\end{figure}
%%%%%%%%%%%%%%%%%%%%%%%%%%%%%%%%%%%%%%%%%%%%%%%%%%%%%%%%%%%
%%%%%%%%%%%%%%%%%%%%%%%%%%%%%%%%%%%%%%%%%%%%%%%%%%%%%%%%%%%%%%%%%%%%%%%%%%%%%%%%%%%%%%%%%%%%%
\begin{figure}[!h]
\begin{center}
\includegraphics[width=18.0cm]{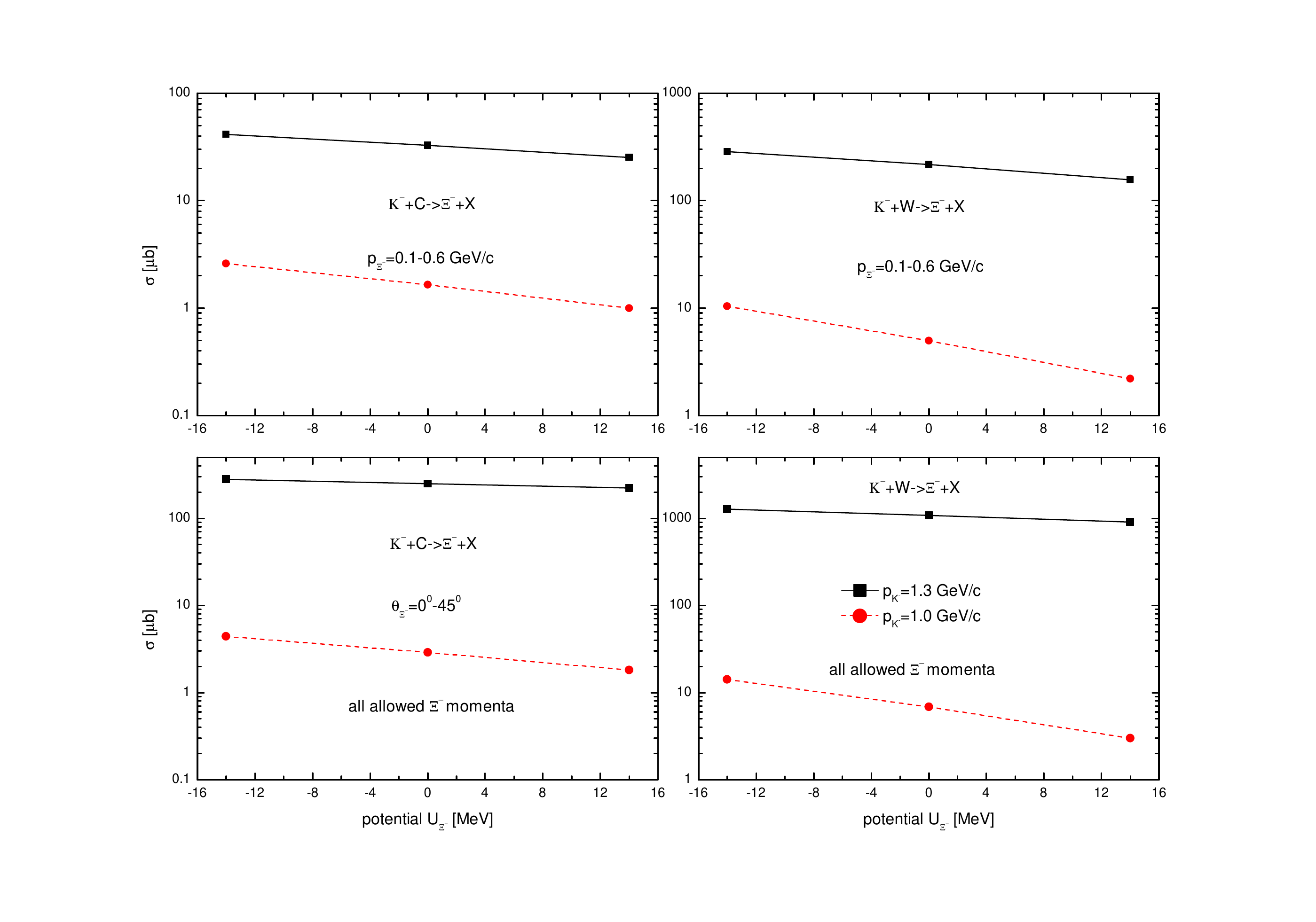}
\vspace*{-2mm} \caption{(Color online) Total cross sections for the production of $\Xi^-$
hyperons from the direct ${K^-}p \to {K^+}{\Xi^-}$ and ${K^-}n \to {K^0}{\Xi^-}$ processes
on $^{12}$C and $^{184}$W nuclei with momenta of 0.1--0.6 GeV/c (upper two panels)
and with all allowed $\Xi^-$ momenta $\ge$ 0.1 GeV/c at given vacuum incident beam momentum
(lower two panels) in the laboratory polar angular range of 0$^{\circ}$--45$^{\circ}$
by medium-modified $K^-$ mesons having vacuum momenta of 1.0 and 1.3 GeV/c,
calculated for the nominal $\Xi^-$ absorption in the nuclear matter,
as functions of the effective scalar $\Xi^-$ potential $U_{\Xi^-}$
at normal nuclear density. The lines are visual guides.}
\label{void}
\end{center}
\end{figure}
%%%%%%%%%%%%%%%%%%%%%%%%%%%%%%%%%%%%%%%%%%%%%%%%%%%%%%%%%%%%%%%%%%%%%%%%%%%%%%%%%%%%%%%%%%
%%%%%%%%%%%%%%%%%%%%%%%%%%%%%%%%%%%%%%%%%%%%%%%%%%%%%%%%%%%%%%%%%%%%%%%%%%%%%%%%%%%%%%%%%%
\begin{figure}[!h]
\begin{center}
\includegraphics[width=18.0cm]{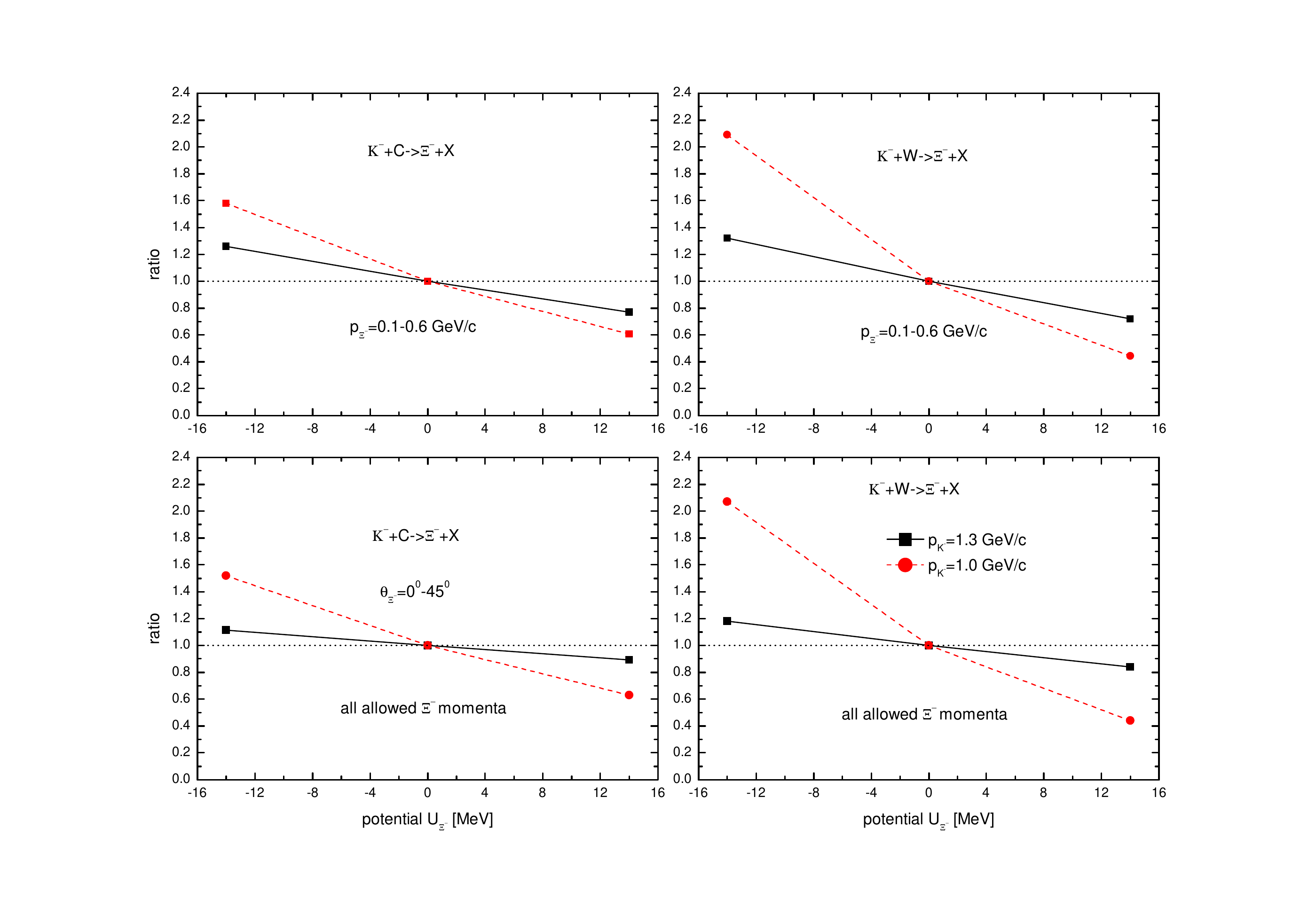}
\vspace*{-2mm} \caption{(Color online) Ratio between the total cross sections for the production of $\Xi^-$
hyperons from the direct ${K^-}p \to {K^+}{\Xi^-}$ and ${K^-}n \to {K^0}{\Xi^-}$
processes on $^{12}$C and $^{184}$W nuclei at laboratory angles of 0$^{\circ}$--45$^{\circ}$ with
momenta of 0.1--0.6 GeV/c (upper two panels) and with all allowed $\Xi^-$ momenta $\ge$ 0.1 GeV/c at given
vacuum incident beam momentum (lower two panels) by medium-modified $K^-$ mesons having vacuum momenta
of 1.0 and 1.3 GeV/c, calculated for the nominal $\Xi^-$ absorption in the nuclear matter
with and without the $\Xi^-$ effective scalar potential $U_{\Xi^-}$ at normal nuclear density,
as function of this potential. The lines are visual guides.}
\label{void}
\end{center}
\end{figure}
%%%%%%%%%%%%%%%%%%%%%%%%%%%%%%%%%%%%%%%%%%%%%%%%%%%%%%%%%%%%%%%%%%%%%%%%%%%%%%%%%%%%%%%%%%%%%
%%%%%%%%%%%%%%%%%%%%%%%%%%%%%%%%%%%%%%%%%%%%%%%%%%%%%%%%%%%%%%%%%%%%%%%%%%%%%%%%%%%%%%%%%%%%%
\begin{figure}[!h]
\begin{center}
\includegraphics[width=18.0cm]{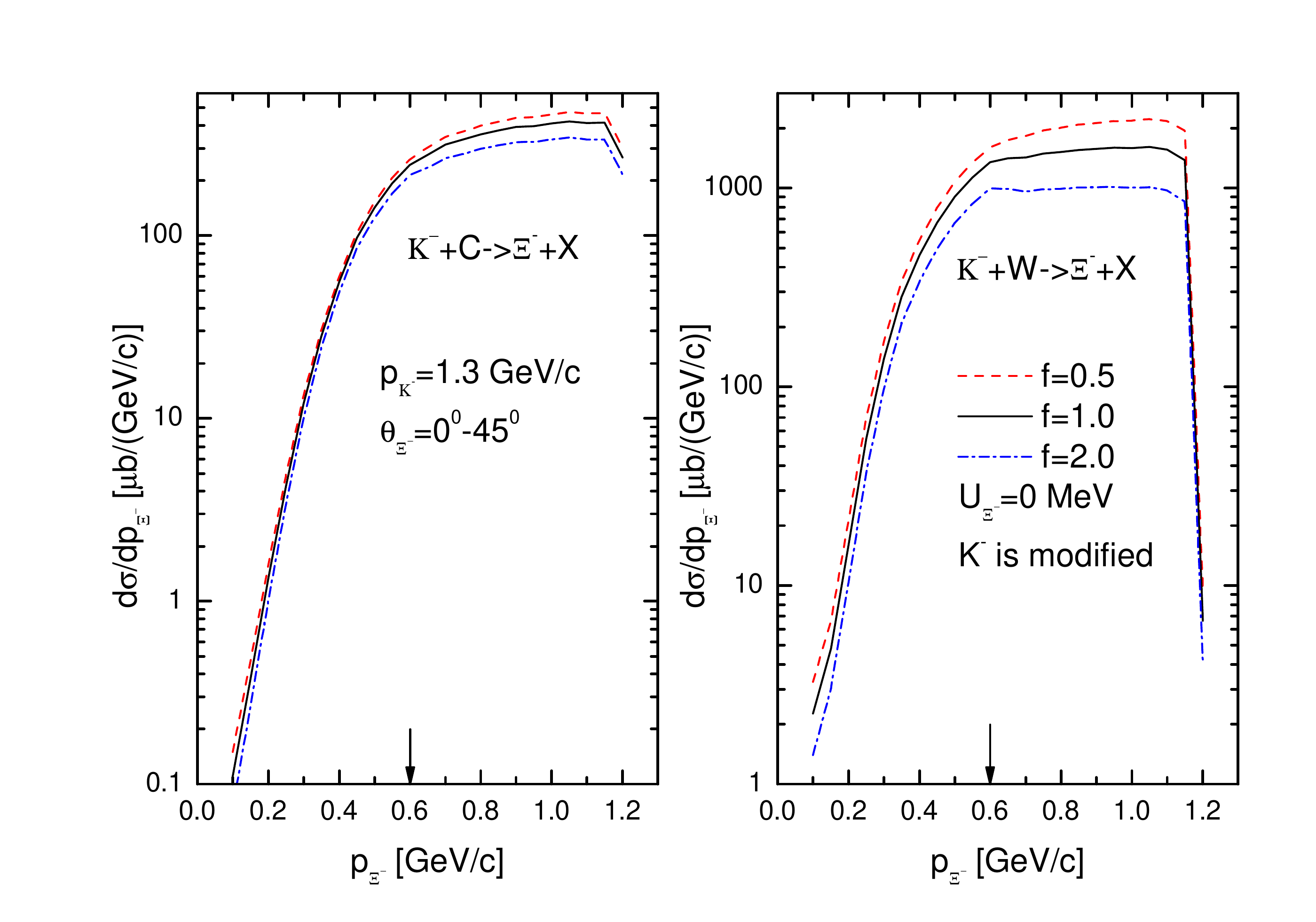}
\vspace*{-2mm} \caption{(Color online) Momentum differential cross sections for the production of $\Xi^-$
hyperons from the direct ${K^-}p \to {K^+}{\Xi^-}$ and ${K^-}n \to {K^0}{\Xi^-}$
processes in the laboratory polar angular range of 0$^{\circ}$--45$^{\circ}$ in the interaction of medium-modified
$K^-$ mesons having vacuum momentum of 1.3 GeV/c with $^{12}$C
(left) and $^{184}$W (right) nuclei, calculated for value of the $\Xi^-$ hyperon effective
scalar potential $U_{\Xi^-}=0$ MeV at density $\rho_0$ assuming that its total inelastic
cross section $\sigma_{{\Xi^-}N}^{\rm in}$ is multiplied by the factors indicated in the inset.
The arrows indicate the boundary between the low-momentum and high-momentum parts of the $\Xi^-$ spectra.}
\label{void}
\end{center}
\end{figure}
%%%%%%%%%%%%%%%%%%%%%%%%%%%%%%%%%%%%%%%%%%%%%%%%%%%%%%%%%%%
\begin{figure}[!h]
\begin{center}
\includegraphics[width=18.0cm]{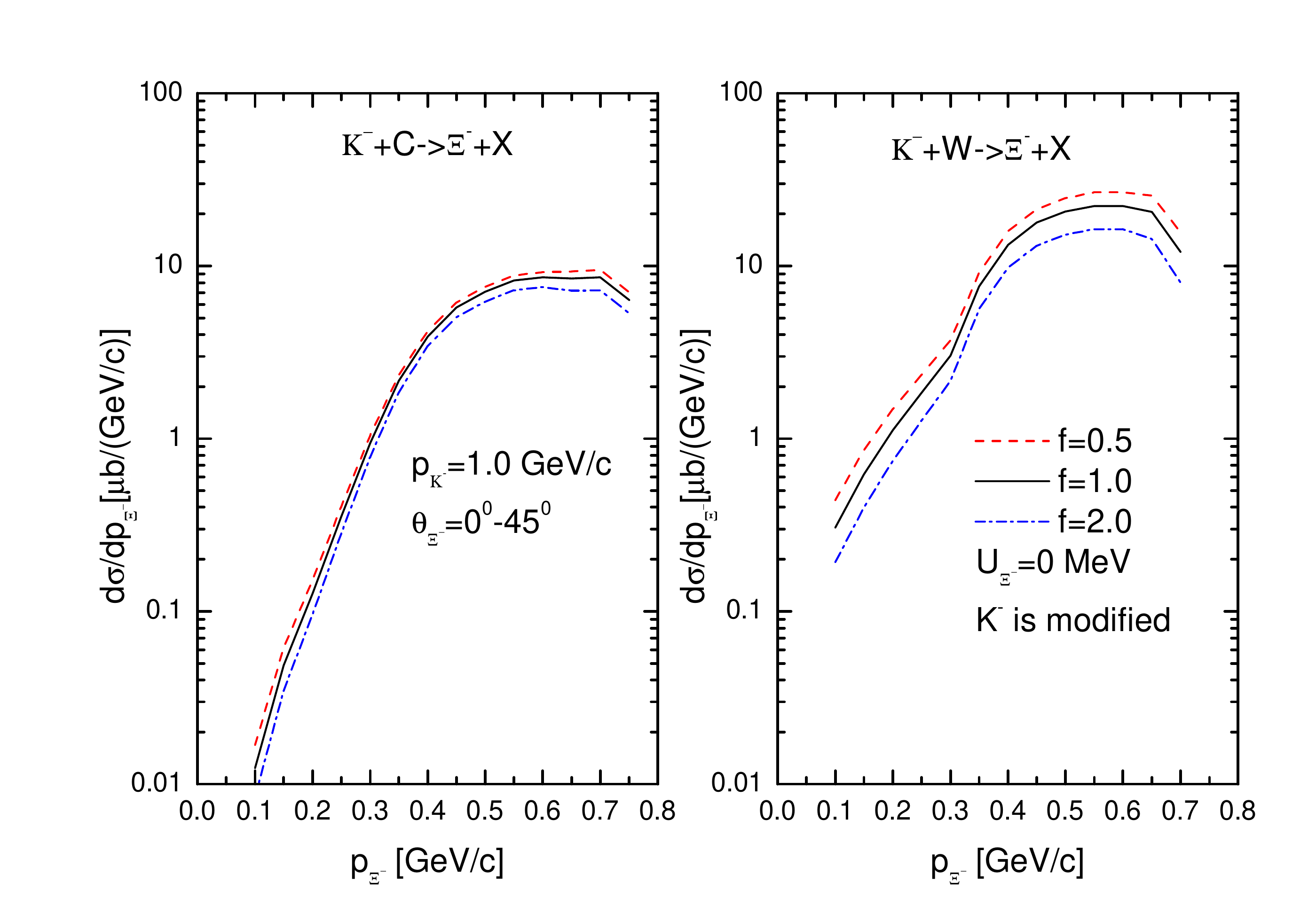}
\vspace*{-2mm} \caption{(Color online) The same as in Fig. 13,
but for the initial vacuum antikaon momentum of 1.0 GeV/c.}
\label{void}
\end{center}
\end{figure}
%%%%%%%%%%%%%%%%%%%%%%%%%%%%%%%%%%%%%%%%%%%%%%%%%%%%%%%%%%%
\begin{figure}[!h]
\begin{center}
\includegraphics[width=16.0cm]{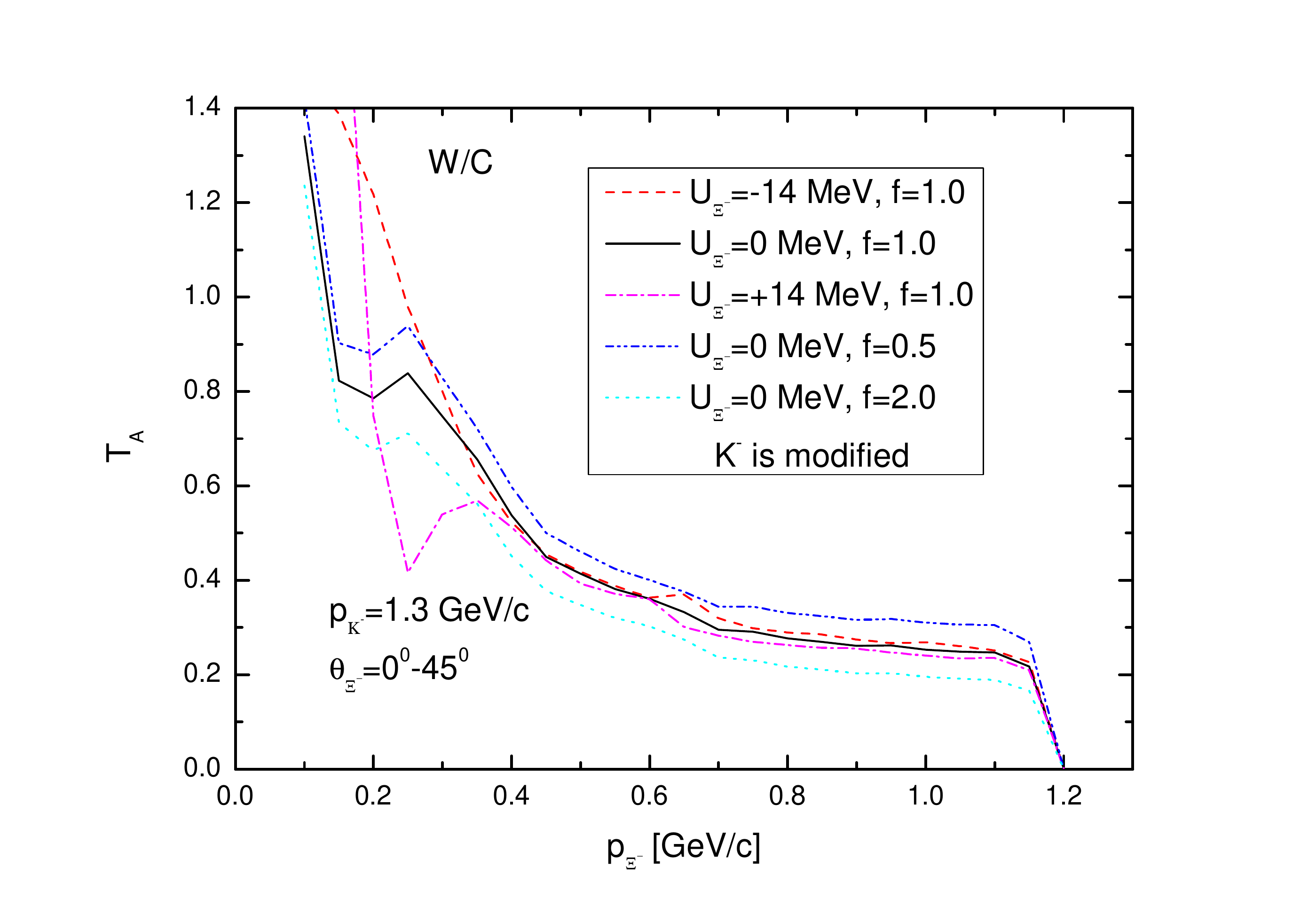}
\vspace*{-2mm} \caption{(Color online) Transparency ratio $T_A$ as a function of the $\Xi^-$  hyperon momentum
for combination $^{184}$W/$^{12}$C as well as for the $\Xi^-$ laboratory polar angular range of
0$^{\circ}$--45$^{\circ}$, for an incident vacuum $K^-$ meson momentum of 1.3 GeV/c and for
different values of its effective scalar potential $U_{\Xi^-}$ at density $\rho_0$ and of the factor $f$
indicated in the inset.}
\label{void}
\end{center}
\end{figure}
%%%%%%%%%%%%%%%%%%%%%%%%%%%%%%%%%%%%%%%%%%%%%%%%%%%%%%%%%%%%%%%%%%%%%%%%%%%%%%%%%%%%%%%%%%%%%
%%%%%%%%%%%%%%%%%%%%%%%%%%%%%%%%%%%%%%%%%%%%%%%%%%%%%%%%%%%
\begin{figure}[!h]
\begin{center}
\includegraphics[width=18.0cm]{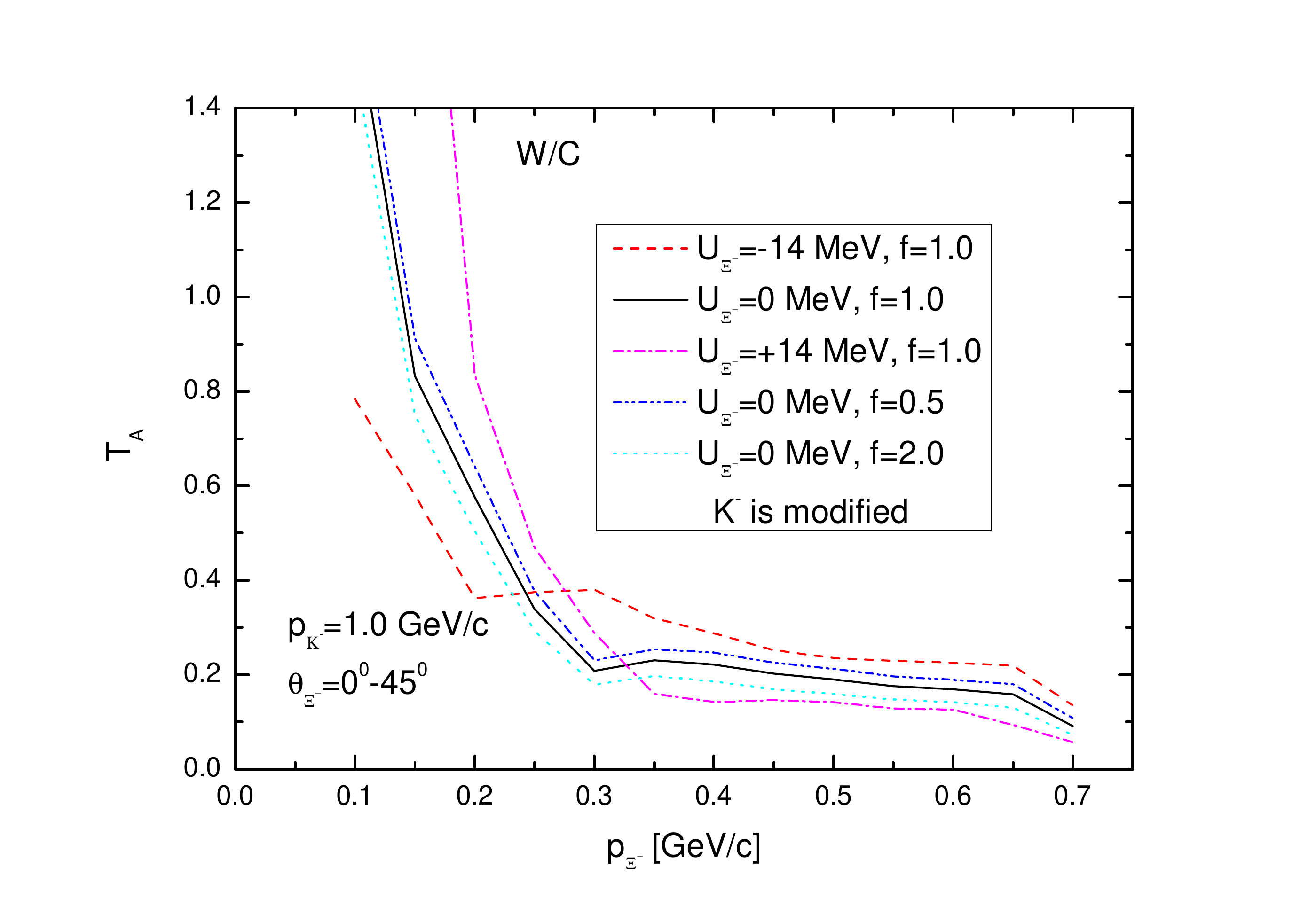}
\vspace*{-2mm} \caption{(Color online) The same as in Fig. 15,
but for the initial vacuum antikaon momentum of 1.0 GeV/c.}
\label{void}
\end{center}
\end{figure}
%%%%%%%%%%%%%%%%%%%%%%%%%%%%%%%%%%%%%%%%%%%%%%%%%%%%%%%%%%%

\section*{3 Results and discussion}

\hspace{0.5cm} At first, we consider the momentum dependences of the absolute $\Xi^-$ hyperon differential
cross sections from the direct production processes (1) and (2) in $K^-$$^{12}$C and $K^-$$^{184}$W interactions.
They were calculated on the basis of Eq. (27) for three basic
adopted values of the $\Xi^-$ effective scalar potential $U_{\Xi^-}$ at density $\rho_0$ for laboratory angles
of 0$^{\circ}$--45$^{\circ}$ and for initial vacuum antikaon momenta of 1.3 and 1.0 GeV/c. Also the nominal
absorption of $\Xi^-$ hyperons in the nuclear matter was accounted for. These dependences are depicted,
respectively, in Figs. 5 and 6. It is seen from these figures
that the $\Xi^-$ hyperon differential cross sections reveal a certain sensitivity to this potential,
mostly in the low-momentum region of 0.1--0.6 GeV/c, for both target nuclei
and for both considered antikaon momenta. Here, the differences between calculations corresponding to
different choices for the $\Xi^-$ scalar potential $U_{\Xi^-}$ are well separated and experimentally
distinguishable. They are practically similar to each other for each target nucleus at initial antikaon momenta
considered. Thus, for example, for incident vacuum $K^-$ meson momentum of 1.3 GeV/c and for outgoing
$\Xi^-$ hyperon momenta of 0.2, 0.4, 0.6 GeV/c the inclusion of the
$\Xi^-$ attractive potential of -14 MeV at normal nuclear matter density leads in the case of $^{12}$C nucleus
to an enhancement of the
$\Xi^-$ production cross sections by factors of about 2.9, 1.4, 1.2, respectively,
as compared to those obtained for the potential $U_{\Xi^-}=0$ MeV.
In the case of $^{184}$W target nucleus these enhancement factors are about 4.5, 1.4, 1.2.
At initial vacuum $K^-$ momentum of 1.0 GeV/c and the same outgoing hyperon momenta of 0.2, 0.4, 0.6 GeV/c
the corresponding enhancement factors are similar and are about 3.1, 1.7, 1.5 and 2.0, 2.2, 2.0 in the cases
of $^{12}$C and $^{184}$W target nuclei, correspondingly. On the other hand, the inclusion of the $\Xi^-$
repulsive potential of +14 MeV at density $\rho_0$ results in the reduction of the $\Xi^-$ hyperon production
cross sections by factors of about 3.5, 1.5, 1.2 and 3.7, 1.6, 1.2 compared to those calculated at zero $\Xi^-$
potential for incoming vacuum antikaon momentum of 1.3 GeV/c and for the same outgoing vacuum $\Xi^-$ momenta
of 0.2, 0.4, 0.6 GeV/c in the cases of $^{12}$C and $^{184}$W target nuclei, respectively.
And for incident beam momentum of 1.0 GeV/c and for the same final $\Xi^-$ momenta these reduction factors
are similar and are about 3.5, 1.9, 1.6 and 2.4, 2.9, 2.1 for $^{12}$C and $^{184}$W, correspondingly.
However, although the $\Xi^-$ hyperon production differential cross sections at beam momentum of 1.0 GeV/c
are less than those at the momentum of 1.3 GeV/c by about of one to two orders of magnitude their strength
at the former momentum is still large enough to be measured in the present experimental facilities.
Thus, the $\Xi^-$ hyperon differential
cross sections measurements in the near-threshold incident $K^-$ momentum region (at 1.0--1.3 GeV/c) will open
a possibility to shed light on its nuclear potential in cold nuclear matter.
Such measurements could be performed in the future at the J-PARC Hadron Experimental Facility
using the high-intensity separated secondary $K^-$ beams. The calculations considered above have been carried out,
supposing that the incoming vacuum $K^-$ meson momentum is modified in the interior of the nucleus in line with
Eqs. (3), (10), (11) due to the presence here of the nuclear attractive optical potential $U_{K^-}$ and the
Coulomb potential $V_{cK^-}$. In order to study the sensitivity of the cascade hyperon production cross sections
from the one-step processes (1), (2) to these potentials, we have used in our calculations also zero values for them.
The results of such calculations for $\Xi^-$ potential $U_{\Xi^-}=0$ MeV at saturation density $\rho_0$
are presented in Figs. 5 and 6 by the dotted curves. One can see that the influence
of the antikaon--nucleon strong and electromagnetic interactions on the free $\Xi^-$ production is negligible at
incident vacuum $K^-$ momentum of 1.3 GeV/c, but favorable for the lower incoming momentum of 1.0 GeV/c.
The latter gives us the need to account for these interactions in our subsequent calculations.

To see more clearly the sensitivity of the differential cross sections, presented in Figs. 5 and 6,
to the $\Xi^-$ hyperon scalar potential $U_{\Xi^-}$ at the central density $\rho_0$, we show in Figs. 7 and 8
the momentum dependences of the ratios of these cross sections, calculated for
the $\Xi^-$ potential $U_{\Xi^-}$, to the analogous cross section, determined at $U_{\Xi^-}=0$ MeV, on a linear scale for $^{12}$C and $^{184}$W nuclei at incident vacuum $K^-$ momenta of 1.3 and 1.0 GeV/c, respectively.
It should be noted that such relative observables are more favorable compared to those based on the
absolute cross sections for the aim of getting the information on particle nuclear potential, since
the theoretical uncertainties associated with the particle production and absorption mechanisms essentially
cancel out in them. It is clearly seen from Fig. 7 that at the incident vacuum $K^-$ meson momentum of 1.3 GeV/c
there are indeed experimentally distinguishable differences at the $\Xi^-$ hyperon momenta $\le$ 0.6 GeV/c
between the results corresponding to the considered options for its scalar potential $U_{\Xi^-}$ for both
target nuclei. At initial subthreshold antikaon momentum of 1.0 GeV/c such differences exist
for these nuclei, as follows from Fig. 8,
in the whole $\Xi^-$ momentum range studied -- at momenta both below and above 0.6 GeV/c.
This means that the in-medium properties of $\Xi^-$ hyperons could be investigated at J-PARC,
using K1.8 or K1.8BR beamlines, through the momentum dependence of their absolute (and relative) production cross sections in inclusive $K^-A$ reactions at initial $K^-$ momenta $\sim$ 1.0--1.3 GeV/c.
It should be pointed out that the ratios of differential cross sections for $\Xi^-$ hyperon production
on $^{184}$W nucleus by 1.0 GeV/c antikaons, presented in Fig. 8, are peaked at $\Xi^-$
momentum $\sim$ 0.3 GeV/c for adopted values of $\pm$14 MeV for its scalar potential $U_{\Xi^-}$ at density $\rho_0$.
This can be explained by the fact that the $\Xi^-$ production cross section, determined at zero value of this
potential, "is bent down" at momenta around this momentum (see Fig. 6) due to the off-shell kinematics
of the direct $K^-N$ collisions and the role played by the nucleus-related effects such as the target proton
binding and Fermi motion, encoded in the nuclear spectral function $P_A({\bf p}_t,E)$. The spectral functions
for nuclei $^{12}$C and $^{184}$W, employed in the present work, are different [75, 103, 104].

   For completeness, in Figs. 9 and 10 we show, respectively, the $\Xi^-$ differential cross sections and
their ratios, analogous to those presented in Figs. 5 and 7, but calculated for the three additional low-energy
$\Xi^-$ potential values of (-25,-15,-5) MeV at density $\rho_0$. It is seen that and in this case the
low-momentum (0.1--0.6 GeV/c) region reveals a definite sensitivity to the $\Xi^-$ potential, which can be
exploited to discriminate experimentally between also these low-energy scenarios for the in-medium
$\Xi^-$ hyperon modification.

The sensitivity of the $\Xi^-$ hyperon production differential cross sections to
its effective scalar potential $U_{\Xi^-}$, demonstrated in Figs. 5, 6 and 7, 8, can also be studied from
such integral measurements as the measurements of the total cross sections for $\Xi^-$ production
in ${K^-}^{12}$C and  ${K^-}^{184}$W interactions at laboratory angles $\le$ 45$^{\circ}$
for the near-threshold incident $K^-$ momenta of 1.0 and 1.3 GeV/c.
Such total cross sections, calculated by integrating Eq. (27) over the $\Xi^-$ momentum $p_{\Xi^-}$
in the low-momentum region (0.1--0.6 GeV/c) and in the full-momentum region allowed for
the given beam momentum are shown in Fig. 11 as functions of this potential.
It can be seen that the low-momentum region of 0.1--0.6 GeV/c shows higher sensitivity
to the potential $U_{\Xi^-}$ than the full-momentum one.
Thus, for instance, the ratios between the total cross sections for $\Xi^-$ hyperon production by
1.0, 1.3 GeV/c $K^-$ mesons on $^{12}$C and $^{184}$W nuclei in this momentum region, calculated with
the potential $U_{\Xi^-}=-14$ MeV, and the cross sections, obtained with $U_{\Xi^-}=+14$ MeV, are
about 2.6, 1.6 and 4.7, 1.8, respectively.
Whereas the same ratios in the full-momentum regions are only somewhat smaller: they are about
2.4, 1.3 for $^{12}$C and 4.7, 1.4 for $^{184}$W, correspondingly.
The highest sensitivity of the total cross sections for $\Xi^-$ production
in the low-momentum and in the full-momentum regions
to the potential $U_{\Xi^-}$ is observed at $K^-$ momentum of 1.0 GeV/c.
Despite the fact that the cross sections at this momentum are smaller than those at beam
momentum of 1.3 GeV/c by about of one -- two orders of magnitude, they have a measurable strength
$\sim$ 1--10 ${\rm \mu}$b. Therefore, the total cross section measurements of $\Xi^-$ hyperon
production on nuclei both in the low-momentum (0.1--0.6 GeV/c) and in the full-momentum regions for
incident antikaon momenta not far below and above threshold
(for momenta $\sim$ 1.0--1.3 GeV/c) will also allow to shed light on its in-medium properties.

Fig. 12 shows the results, which also support
the findings of Fig. 11 that the total $\Xi^-$ hyperon production cross sections
both in the low-momentum (0.1--0.6 GeV/c) and in the full-momentum regions reveal
a some sensitivity to its in-medium scalar potential $U_{\Xi^-}$ at saturation density $\rho_0$
for the considered incident $K^-$ momenta.
Here, the ratios of the $\Xi^-$ production total cross sections, calculated for the
potential $U_{\Xi^-}$ and presented in Fig. 11, to the analogous cross sections,
determined at $U_{\Xi^-}=0$ MeV, are shown as functions of this potential.
It is nicely seen that the highest sensitivity of the ratios in
both considered kinematic regions to the potential $U_{\Xi^-}$ is indeed observed at
subthreshold $K^-$ momentum of 1.0 GeV/c.
Thus, at this momentum and for these regions the cross section ratios for $U_{\Xi^-}=-14$ MeV
are about 1.5 and 2.1 for $^{12}$C and $^{184}$W targets, respectively.
As the antikaon momentum increases to 1.3 GeV/c, the sensitivity of the total cross section
ratios to variations in the $\Xi^-$ hyperon effective scalar potential $U_{\Xi^-}$ decreases.
Thus, in the case where $\Xi^-$ hyperons of
momenta of 0.1--0.6 GeV/c are produced by 1.3 GeV/c $K^-$ mesons impinging on $^{12}$C and $^{184}$W
nuclei, the considered ratios for $U_{\Xi^-}=-14$ MeV take smaller but yet a measurable values of about 1.3
and 1.3, respectively. The analogous ratios for the production of the $\Xi^-$ hyperons
in the full-momentum regions by 1.3 GeV/c antikaons in $^{12}$C and $^{184}$W target nuclei are
yet smaller: they are $\sim$ 1.1 and 1.2, respectively.

To study the sensitivity of the $\Xi^-$ hyperon production cross sections to its absorption
cross section in nuclear medium in near-threshold antikaon-induced reactions, we calculated
the absolute $\Xi^-$ momentum differential cross sections from the direct processes (1) and (2) in
${K^-}^{12}$C and ${K^-}^{184}$W collisions for an incident vacuum $K^-$ momenta of 1.3 and 1.0 GeV/c
at laboratory angles $\le$ 45$^{\circ}$ for zero value of its potential $U_{\Xi^-}$ at density $\rho_0$
assuming that ${\Xi^-}N$ nominal inelastic cross section, given by Eqs. (17), (31) and (32),
is multiplied by the factor $f=0.5$, 1.0, 2.0. They are shown in Figs. 13 and 14, respectively.
It is seen from Fig. 13 that  there are a sizeable
differences ($\sim$ 10--20\% for $^{12}$C and $\sim$ 20--40\% for $^{184}$W) between all calculations
corresponding to different considered choices for the ${\Xi^-}N$ inelastic cross section. While these
differences are comparable in the high-momentum range of 0.6--1.2 GeV/c
(where the $\Xi^-$ production cross sections are the greatest)
with those caused by the modification of the $\Xi^-$ hyperon mass in nuclear
matter due to the strong interaction, they are significantly less than the latter ones at
momenta below 0.6 GeV/c (cf. Fig. 5), where the role of the $\Xi^-$ scalar potential is essential.
At initial $\Xi^-$ momentum of 1.0 GeV/c the latter differences
are significantly larger, as follows from Figs. 6 and 14, than the former ones at all final $\Xi^-$ momenta.
This leads to the important conclusion that a comparison of the "differential" and "integral"
results depicted in Figs. 5--12 with the corresponding experimental data, which could be taken
in the dedicated experiment at J-PARC, should allow one to distinguish at least
between zero, possible weak attractive ($U_{\Xi^-}$ $\sim$ -14 MeV) and weak repulsive
($U_{\Xi^-}$ $\sim$ +14 MeV) $\Xi^-$ hyperon scalar potentials in cold nuclear matter
in spite of the fact that the ${\Xi^-}N$ inelastic cross section is experimentally unknown at low
$\Xi^-$ momenta.

Finally, it is also interesting to clarify additionally the opportunity of extracting the strength
of the ${\Xi^-}$ hyperon effective scalar potential at saturation density $\rho_0$
from the measurements of such relative observable as the transparency ratio $T_A$ for $\Xi^-$ hyperons,
as this has been done for the $\eta^{\prime}$ and $J/\psi$ mesons in Refs. [114] and [115], respectively.
Figures 15 and 16 show the momentum dependence of this quantity for the W/C combination
for $\Xi^-$ hyperons produced in the primary channels (1) and (2) at laboratory angles $\le$ 45$^{\circ}$
by 1.3 and 1.0 GeV/c antikaons, respectively. It is calculated according to Eq. (28)
for the adopted values of the $\Xi^-$  hyperon effective scalar potential $U_{\Xi^-}$ at density $\rho_0$
and of the factor $f$, by which its total inelastic cross section is multiplied.
One can see that for the subthreshold $K^-$ meson momentum of 1.0 GeV/c there is, contrary to the case of
its above threshold momentum of 1.3 GeV/c, a strong sensitivity ($\sim$ 40--60\%) of the
transparency ratio $T_A$ to the considered variations in the nuclear potential $U_{\Xi^-}$ at all outgoing
$\Xi^-$ momenta, which cannot be practically masked by that ($\sim$ 10--20\%) associated with
the possible changes in the ${\Xi^-}N$ inelastic cross section. This means that the future precise
$\Xi^-$ production data on the momentum dependence of the transparency ratio $T_A$ for $\Xi^-$ hyperons,
taken for the initial antikaon momentum of 1.0 GeV/c, should also help, with accounting for
the results given in Figs. 15 and 16, to determine the $\Xi^-$ hyperon effective scalar potential
in cold nuclear matter at its density $\rho_0$.

Thus, taking into account the above considerations, one can conclude that the
$\Xi^-$ differential and total cross section measurements in antikaon--nucleus reactions at initial
momenta not far from threshold (at momenta $\sim$ 1.0--1.3 GeV/c) as well as the transparency ratio
measurements for $\Xi^-$ hyperons at subthreshold momenta $\sim$ 1.0 GeV/c
will allow us to shed light on this potential.

\section*{4 Summary}

\hspace{0.5cm} In the present paper we study the antikaon-induced inclusive cascade $\Xi^-$ hyperon
production from $^{12}$C and $^{184}$W target nuclei near threshold within a nuclear spectral function
approach. The approach describes incoherent direct $\Xi^-$ hyperon production in elementary
${K^-}p \to {K^+}\Xi^-$ and ${K^-}n \to {K^0}\Xi^-$ processes as well as takes into account the influence
of the scalar nuclear $K^-$, $K^+$, $K^0$, $\Xi^-$ and their Coulomb potentials on these processes.
We calculate the absolute differential and total cross sections for the production of $\Xi^-$ hyperons
off these nuclei at laboratory angles $\le$ 45$^{\circ}$ by $K^-$ mesons with momenta of 1.0 and 1.3 GeV/c,
which are close to the threshold momentum (1.05 GeV/c) for $\Xi^-$ hyperon production on the free
target nucleon at rest. We also calculate the momentum dependence of the transparency ratio for the
$^{184}$W/$^{12}$C combination for $\Xi^-$ hyperons at these $K^-$ beam momenta.
We show that the $\Xi^-$ differential and total (absolute and relative) production cross sections
at the considered initial momenta reveal a distinct sensitivity to the variations
in the scalar $\Xi^-$ nuclear potential at saturation density $\rho_0$, studied in the paper,
in the low-momentum region of 0.1--0.6 GeV/c.
We also demonstrate that for the subthreshold $K^-$ meson momentum of 1.0 GeV/c there is, contrary to the
case of its above threshold momentum of 1.3 GeV/c, a strong sensitivity of the transparency ratio
for $\Xi^-$ hyperons to the considered changes in the $\Xi^-$ nuclear potential at all outgoing $\Xi^-$
momenta as well, which cannot be masked, as in the case of differential and total observables mentioned above,
by that associated with the possible changes in the poorly known experimentally ${\Xi^-}N$ inelastic cross section. Therefore, the measurement of these absolute and relative observables in a dedicated experiment at
the J-PARC Hadron Experimental Facility will provide valuable information on the $\Xi^-$ in-medium properties,
which will be complementary to that deduced from the study of the inclusive ($K^-$,$K^+$) reactions
at incident momenta of 1.6--1.8 GeV/c in the $\Xi^-$ bound and quasi-free regions.
\\
\\

%%%%%%%%%%%%%%%%%%%%%%%%%%%%%%%%%%%%%%%%%%%%%%%%%%%%%%%%%%%%%%%%

\begin{thebibliography}{99}
\bibitem{1} W. Cassing and E. L. Bratkovskaya, Phys. Rep. {\bf 308}, 65 (1999).\\
                  C. Fuchs. Prog. Part. Nucl. Phys. {\bf 56}, 1 (2006);\\
                                  arXiv:nucl-th/0507017.\\                                                                              C. Hartnack {\it et al.}, Phys. Rep. {\bf 510}, 119 (2012);\\
                                    arXiv:1106.2083 [nucl-th].\\
                        O. Buss {\it et al.}, Phys. Rep. {\bf 512}, 1 (2012);\\
                                    arXiv:1106.1344 [hep-ph].
\bibitem{2} E. Friedman and A. Gal, Phys. Rep. {\bf 452}, 89 (2007);\\
                                    arXiv:0705.3965 [nucl-th].
\bibitem{3} L. Tolos and L. Fabbietti, Prog. Part. Nucl. Phys. {\bf 112}, 103770 (2020);\\
                            arXiv:2002.09223 [nucl-ex].
\bibitem{4} L. Tolos, R. Molina, E. Oset, and A. Ramos, Phys. Rev. C {\bf 82}, 045210 (2010);\\
                                        arXiv:1006.3454 [nucl-th].
\bibitem{5} E. Oset {\it et al.}, Int. J. Mod. Phys. E {\bf 21}, 1230011 (2012);\\
                                        arXiv:1210.3738 [nucl-th].
\bibitem{6} A. Ilner, D. Cabrera, P. Srisawad, and E. Bratkovskaya, Nucl. Phys. A {\bf 927}, 249 (2014);\\
                                        arXiv:1312.5215 [hep-ph].
\bibitem{7} D. Cabrera {\it et al.}, Journal of Physics: Conf. Series {\bf 503}, 012017 (2014);\\
                                        arXiv:1312.4343 [hep-ph].\\
             L. Tolos, EPJ Web of Conf. {\bf 171}, 09003 (2018).
\bibitem{8} K. Tsushima, A. Sibirtsev, and A. W. Thomas, Phys. Rev. C {\bf 62}, 064904 (2000);\\
                                        arXiv:nucl-th/0004011.
\bibitem{9} D. Suenaga and P. Lakaschus, Phys. Rev. C {\bf 101}, 035209 (2020);\\
                                        arXiv:1908.10509 [nucl-th].
\bibitem{10} E. Ya. Paryev, Chinese Physics C, Vol. {\bf 44}, No. (11), 114106 (2020);\\
                             arXiv:2007.10192 [nucl-th].
\bibitem{11} S. H. Lee and S. Cho, Int. J. Mod. Phys. E {\bf 22}, 1330008 (2013);\\
                            arXiv:1302.0642 [nucl-th].
\bibitem{12} T. Song, T. Hatsuda, and S. H. Lee, Phys. Lett. B {\bf 792}, 160 (2019);\\
                            arXiv:1808.05372 [nucl-th].
\bibitem{13} S. H. Lee, arXiv:1904.09064 [nucl-th].
\bibitem{14} S. H. Lee, Nucl. Part. Phys. Proc. {\bf 309--311}, 111 (2020).
\bibitem{15} E. Ya. Paryev, Nucl. Phys. A {\bf 1007}, 122133 (2021);\\
                            arXiv:2102.00789 [nucl-th].
\bibitem{16} M. M. Kaskulov and E. Oset, Phys. Rev. C {\bf 73}, 045213 (2006);\\
                            arXiv:nucl-th/0509088.
\bibitem{17} M. M. Kaskulov and E. Oset, AIP Conf. Proc. {\bf 842}, 483--5 (2006).
\bibitem{18} M. F. M. Lutz, C. L. Copra and M. Moeller, Nucl. Phys. A {\bf 808}, 124 (2008);\\
                            arXiv:0707.1283 [nucl-th].
\bibitem{19} D. Cabrera {\it et al.}, Phys. Rev. C {\bf 90}, 055207 (2014);\\
                            arXiv:1406.2570 [hep-ph].
\bibitem{20} S. Petschauer {\it et al.}, arXiv:2002.00424 [nucl-th].
\bibitem{21} Z. Q. Feng, W. J. Xie, and G. M. Jin, Phys. Rev. C {\bf 90}, 064604 (2014).
\bibitem{22} E. Ya. Paryev and Yu. T. Kiselev, Nucl. Phys. A {\bf 992}, 121622 (2019);\\
                            arXiv:1910.02755 [nucl-th].
\bibitem{23} M. Kaskulov, L. Roca and E. Oset, Eur. Phys. J. A {\bf 28}, 139 (2006);\\
                            arXiv:nucl-th/0601074.
\bibitem{24} E. Ya. Paryev, Phys. Atom. Nucl. Vol. {\bf 75}, No.12, 1523 (2012).
\bibitem{25} E. Ya. Paryev, J. Phys. G: Nucl. Part. Phys. {\bf 37}, 105101 (2010);\\
                            arXiv:1010.0111 [nucl-th].
\bibitem{26} A. Gal, E. V. Hungerford and D. J. Millener, Rev. Mod. Phys. {\bf 88}, 035004 (2016);\\                                    arXiv:1605.00557 [nucl-th].
\bibitem{27} T. Hatsuda {\it et al.}, Nucl. Phys. A {\bf 967}, 856 (2017);\\
                            arXiv:1704.05225 [nucl-th].
\bibitem{28} K. Sasaki {\it et al.} (HAL QCD Collaboration), Nucl. Phys. A {\bf 998}, 121737 (2020);\\
                            arXiv:1912.08630 [hep-lat].\\
             K. Sasaki {\it et al.} (HAL QCD Collaboration), EPJ Web Conf. {\bf 175}, 05010 (2018).
\bibitem{29} C. B. Dover and A. Gal, Annals of Phys. {\bf 146}, 309 (1983).
\bibitem{30} S. Aoki {\it et al.}, Phys. Lett. B {\bf 355}, 45 (1995).
\bibitem{31} T. Fukuda {\it et al.}, Phys. Rev. C {\bf 58}, 1306 (1998).
\bibitem{32} P. Khaustov {\it et al.}, Phys. Rev. C {\bf 61}, 054603 (2000);\\
                                      arXiv:nucl-ex/9912007.
\bibitem{33} T. Iijima {\it et al.}, Nucl. Phys. A {\bf 546}, 588 (1992).
\bibitem{34} T. Nagae {\it et al.} (J-PARC E05 Collaboration), PoS INPC {\bf 2016}, 038 (2017);\\
                                     AIP Conf. Proc. {\bf 2130}, 020015 (2019).
\bibitem{35} H. Maekawa {\it et al.}, arXiv:0704.3929 [nucl-th].
\bibitem{36} H. Maekawa, K. Tsubakihara and A. Ohnishi, Eur. Phys. J. A {\bf 33}, 269 (2007);\\
                                      arXiv:nucl-th/0701066.
\bibitem{37} J. Hu and H. Shen, Phys. Rev. C {\bf 96}, 054304 (2017);\\
                                      arXiv:1710.08613 [nucl-th].
\bibitem{38} E. Hiyama {\it et al.}, Phys. Rev. C {\bf 78}, 054316 (2008);\\
                                      arXiv:0811.3156 [nucl-th].
\bibitem{39} Y. Jin, X.-R. Zhou, Yi-Yu. Cheng and H.-J. Schulze, arXiv:1910.05884 [nucl-th].

\bibitem{40} T. Miyatsu and K. Saito,  Prog. Theor. Phys. {\bf 122}, 1035 (2009);\\
                                        arXiv:0903.1893 [nucl-th].
\bibitem{41} M. Kohno and Y. Fujiwara, Phys. Rev. C {\bf 79}, 054318 (2009);\\
                                    arXiv:0904.0517 [nucl-th].
\bibitem{42} K. Nakazawa {\it et al.}, Prog. Theor. Exp. Phys. {\bf 2015}, 033D02 (2015).
\bibitem{43} S. H. Hayakawa {\it et al.} (J-PARC E07 Collaboration), Phys. Rev. Lett. {\bf 126}, 062501 (2021);\\
                            arXiv:2010.14317 [nucl-ex].
\bibitem{44} M. Yoshimoto {\it et al.}, arXiv:2103.08793 [nucl-ex].
\bibitem{45} E. Hiyama {\it et al.}, Phys. Rev. Lett. {\bf 124}, 092501 (2020);\\
                            arXiv:1910.02864 [nucl-th].
\bibitem{46} G. Meher and U. Raha, arXiv:2010.12291 [nucl-th].
\bibitem{47} Md. A. Khan {\it et al.}, arXiv:2011.03708 [nucl-th].
\bibitem{48} H. Ohnishi, F. Sakuma, and T. Takahashi, arXiv:1912.02380 [nucl-ex].
\bibitem{49} B. Abelev {\it et al.} (ALICE Collaboration), Phys. Lett. B {\bf 728}, 216 (2014);\\
                                  arXiv:1307.5543 [nucl-ex].
\bibitem{50} J. Castillo (STAR Collaboration), Nucl. Phys. A {\bf 715}, 518 (2003);\\
                                        arXiv:nucl-ex/0210032.
\bibitem{51} J. Adams {\it et al.} (STAR Collaboration), Phys. Rev. Lett. {\bf 98}, 062301 (2007);\\
                            arXiv:nucl-ex/0606014.
\bibitem{52} M. M. Aggarwal {\it et al.} (STAR Collaboration), Phys. Rev. C {\bf 83}, 024901 (2011);\\
                                      arXiv:1010.0142 [nucl-ex].
\bibitem{53} S. V. Afanasiev {\it et al.} (NA49 Collaboration), Phys. Lett. B {\bf 538}, 275 (2002);\\
                                      arXiv:hep-ex/0202037.
\bibitem{54} F. Antinori {\it et al.} (NA57 Collaboration), Phys. Lett. B {\bf 595}, 68 (2004).\\
             F. Antinori {\it et al.} (NA57 Collaboration), J. Phys. G {\bf 31}, 1345 (2005);\\
                                       arXiv:nucl-ex/0509009.
\bibitem{55} C. Alt {\it et al.} (NA49 Collaboration), Phys. Rev. C {\bf 78}, 034918 (2008);\\
                                      arXiv:0804.3770 [nucl-ex].
\bibitem{56} P. Chung {\it et al.} (E895 Collaboration), Phys. Rev. Lett. {\bf 91}, 202301 (2003);\\
                            arXiv:nucl-ex/0302021.
\bibitem{57} G. Agakishiev {\it et al.} (HADES Collaboration), Phys. Rev. Lett. {\bf 103}, 132301 (2009);\\
                            arXiv:0907.3582 [nucl-ex].
\bibitem{58} A. Andronic, P. Braun-Munzinger, and K. Redlich, Nucl. Phys. A {\bf 765}, 211 (2006).
\bibitem{59} L.-W. Chen, C. M. Ko, and Y. Tzeng, Phys. Lett. B {\bf 584}, 269 (2004);\\
                                      arXiv:nucl-th/0312009.
\bibitem{60} F. Li, L.-W. Chen, C. M. Ko and S. H. Lee, Phys. Rev. C {\bf 85}, 064902 (2012);\\
                                    arXiv:1204.1327 [nucl-th].
\bibitem{61} G. Graef {\it et al.}, Phys. Rev. C {\bf 90}, 064909 (2014);\\
                                      arXiv:1409.7954 [nucl-th].
\bibitem{62} A. Aduszkiewicz {\it et al.} (NA61/SHINE Collaboration), arXiv:2006.02062 [nucl-ex].
\bibitem{63} F. Antinori {\it et al.} (NA57 Collaboration), J. Phys. G {\bf 32}, 427 (2006);\\
                                      arXiv:nucl-ex/0601021.
\bibitem{64} G. Agakishiev {\it et al.} (HADES Collaboration), Phys. Rev. Lett. {\bf 114}, 212301 (2015);\\
                                      arXiv:1501.03894 [nucl-ex].
\bibitem{65} J. W. Price {\it et al.} (CLAS Collaboration), Nucl. Phys. A {\bf 754}, 272c (2005);\\
                                      arXiv:nucl-ex/0402006.
\bibitem{66} J. W. Price {\it et al.} (CLAS Collaboration), Phys. Rev. C {\bf 71}, 058201 (2005);\\
                                      arXiv:nucl-ex/0409030.
\bibitem{67} L. Guo {\it et al.} (CLAS Collaboration), Phys. Rev. C {\bf 76}, 025208 (2007);\\
                                      arXiv:nucl-ex/0702027.
\bibitem{68} G. Barucca {\it et al.} (PANDA Collaboration), arXiv:2009.11582 [hep-ex];\\
                           arXiv:2012.01776 [hep-ex]; arXiv:2101.11877 [hep-ex].
\bibitem{69} J. Adamczewski-Musch {\it et al.} (HADES Collaboration with PANDA@HADES Collaboration),\\
                                      arXiv:2010.06961 [nucl-ex].
\bibitem{70} Y. Nara {\it et al.}, Nucl. Phys. A {\bf 614}, 433 (1997).
\bibitem{71} E. Ya. Paryev, Chinese Physics C, Vol. {\bf 42}, No. (8), 084101 (2018);\\
                             arXiv:1806.00303 [nucl-th].

\bibitem{72} E. Ya. Paryev, M. Hartmann and Yu. T. Kiselev, J. Phys. G:
                      Nucl. Part. Phys. {\bf 42}, 075107 (2015);
                      arXiv:1505.01992 [nucl-th].
\bibitem{73} E. Ya. Paryev, Eur. Phys. J. A {\bf 23}, 453 (2005).
\bibitem{74} M. Kohno, Phys. Rev. C {\bf 100}, 024313 (2019);
                        arXiv:1908.01934 [nucl-th].
\bibitem{75} E. Ya. Paryev, Eur. Phys. J. A {\bf 9}, 521 (2000).
\bibitem{76} V. Metag, M. Nanova, and E. Ya. Paryev, Prog. Part. Nucl. Phys. {\bf 97}, 199 (2017);\\
                            arXiv:1706.09654 [nucl-ex].
\bibitem{77} K. Tsushima {\it et al.}, Phys. Lett. B {\bf 429}, 239 (1998).
\bibitem{78} E. Ya. Paryev, M. Hartmann and Yu. T. Kiselev, Chinese Physics C, \\ Vol.{\bf 41}, No.12, 124108 (2017);
                             arXiv:1612.02767 [nucl-th].
\bibitem{79} A. Sibirtsev and W. Cassing, Nucl. Phys. A {\bf 641}, 476 (1998);
                                        arXiv:nucl-th/9805021.
\bibitem{80} G. Q. Li and C. M. Ko, Phys. Rev. C {\bf 54}, 1897 (1996);
                                        arXiv:nucl-th/9608049.
\bibitem{81} Z. Q. Feng, Phys. Rev. C {\bf 101}, 064601 (2020);
                        arXiv:2006.02247 [nucl-th].
\bibitem{82} C.-H. Lee {\it et al.}, Phys. Lett. B {\bf 412}, 235 (1997);
                                      arXiv:nucl-th/9705012.
\bibitem{83} C. B. Dover and G. E. Walker, Phys. Rep. {\bf 89}, 1 (1982).
\bibitem{84} N.-Y. Ghim {\it et al.}, arXiv:2102.05292 [nucl-th].
\bibitem{85} T. Gaitanos and A. Chorozidou, arXiv:2101.08470 [nucl-th].
\bibitem{86} T. Inoue (for HAL QCD Collaboration), {\it AIP Conf. Proc.} {\bf 2130}, no.1, 020002 (2019);\\
                                       arXiv:1809.08932 [hep-lat].
\bibitem{87} T. Inoue (for HAL QCD Collaboration), {\it PoS} {\bf INPC 2016}, 277 (2016);\\
                                       arXiv:1612.08399 [hep-lat].
\bibitem{88} J. Haidenbauer and  U.- G. Meissner, Eur. Phys. J. A {\bf 55}, 23 (2019);\\
                                       arXiv:1810.04883 [nucl-th].
\bibitem{89} J. Haidenbauer, U.- G. Meissner, and S. Petschauer, Nucl. Phys. A {\bf 954}, 273 (2016);\\                                     arXiv:1511.05859 [nucl-th].
\bibitem{90} M. M. Nagels, Th. A. Rijken, and Y. Yamamoto, arXiv:1504.02634 [nucl-th].
\bibitem{91} M. Kohno, Phys. Rev. C {\bf 81}, 014003 (2010);\\
                                    arXiv:0912.4330 [nucl-th].
\bibitem{92} H. Polinder, J. Haidenbauer, and U.-G. Meissner, Phys. Lett. B {\bf 653}, 29 (2007);\\
                            arXiv:0705.3753 [nucl-th].
\bibitem{93} M. Nanova {\it et al.} (CBELSA/TAPS Collaboration), Phys. Lett. B {\bf 727}, 417 (2013);\\
                            arXiv:1311.0122 [nucl-ex].
\bibitem{94} M. Nanova {\it et al.} (CBELSA/TAPS Collaboration), Phys. Rev. C {\bf 94}, 025205 (2016);\\
                            arXiv:1607.07228 [nucl-ex].
\bibitem{95} T. Harada and Y. Hirabayashi, Phys. Rev. C {\bf 103}, 024605 (2021);\\
                                    arXiv:2101.00855 [nucl-th].
\bibitem{96} E. E. Kolomeitsev, B. Tomasik, and D. N. Voskresensky, Phys. Rev. C {\bf 86}, 054909 (2012);\\
                                    arXiv:1207.5738 [nucl-th].
\bibitem{97} B. Tomasik and E. E. Kolomeitsev, arXiv:1510.04349 [nucl-th].
\bibitem{98} T. Harada and Y. Hirabayashi, Phys. Rev. C {\bf 102}, 024618 (2020);\\
                                    arXiv:2006.15627 [nucl-th].
\bibitem{99} E. Friedman and A. Gal, arXiv:2104.00421 [nucl-th].
\bibitem{100} T. Tamagawa {\it et al.}, Nucl. Phys. A {\bf 691}, 234c (2001).
\bibitem{101} J. K. Ahn {\it et al.}, Phys. Lett. B {\bf 633}, 214 (2006);\\
                                         arXiv:nucl-ex/0502010.
\bibitem{102} J. K. Ahn and S.-il Nam, arXiv:2101.10114 [hep-ph].
\bibitem{103} S. V. Efremov and  E. Ya. Paryev, Eur. Phys. J. A {\bf 1}, 99 (1998).
\bibitem{104} E. Ya. Paryev, Eur. Phys. J. A {\bf 7}, 127 (2000).
\bibitem{105} A. Sibirtsev {\it et al.}, Z. Phys. A {\bf 351}, 333 (1995).
\bibitem{106} A. Sibirtsev and W. Cassing, arXiv:nucl-th/9909053.
\bibitem{107} V. Flaminio {\it et al.}, Compilation of Cross Sections.\\
             II: ${K^+}$ and ${K^-}$ Induced Reactions. CERN-HERA {\bf 83-02}, (1983).
\bibitem{108} D. A. Sharov, V. L. Korotkikh, and  D. E. Lanskoy, Eur. Phys. J. A {\bf 47}, 109 (2011);\\
                              arXiv:1105.0764 [nucl-th].
\bibitem{109} J. K. Ahn {\it et al.}, Nucl. Phys. A {\bf 625}, 231 (1997).
\bibitem{110} S. Aoki {\it et al.}, Nucl. Phys. A {\bf 644}, 365 (1998).
\bibitem{111} S. J. Kim {\it et al.}, presentation at the 12th Int. Conf. on Hypernuclear and\\
             Strange Particle Physics. Sendai, Japan (2015);\\
             http://lambda.phys.tohoku.ac.jp/hyp2015/
\bibitem{112} J. Haidenbauer and U.- G. Meissner, Nucl. Phys. A {\bf 936}, 29 (2015);\\                                     arXiv:1411.3114 [nucl-th].
\bibitem{113} S. Petschauer {\it et al.}, Eur. Phys. J. A {\bf 52}, 15 (2016);\\
                                   arXiv:1507.08808 [nucl-th].
\bibitem{114} E. Ya. Paryev, J. Phys. G: Nucl. Part. Phys. {\bf 40}, 025201 (2013);\\
                                   arXiv:1209.4050 [nucl-th].
\bibitem{115} E. Ya. Paryev, Yu. T. Kiselev, and Yu. M. Zaitsev, Nucl. Phys. A {\bf 968}, 1 (2017).
\end{thebibliography}
\end{document}